%% file: main.tex
\pdfoutput=1
\documentclass[11pt]{article}
\usepackage{graphicx}
\usepackage{textcase}
\usepackage[margin=1.25in]{geometry}
\usepackage[usenames,dvipsnames]{color}
\usepackage{url}
\usepackage{paracol}
\usepackage[breaklinks,
    colorlinks=true,
    urlcolor=blue,
    anchorcolor=blue,
    citecolor=blue,
    filecolor=blue,
    linkcolor=blue,
    menucolor=blue,
    linktocpage=true,
    pdfproducer=medialab,
    pdfa=true
]{hyperref}
\usepackage{authblk} %for author block
\usepackage{enumitem}
\usepackage{pdfpages}
\usepackage{breakurl}
\usepackage[toc,page,title]{appendix}
\usepackage[toc]{glossaries}
\usepackage{setspace}

\usepackage{cleveref}
\AtBeginEnvironment{appendices}{\crefalias{section}{appendix}}

\usepackage{tcolorbox}
\newtcbox{\lead}{on line,
  colframe=orange,colback=yellow,
  boxrule=0.5pt,arc=2pt,boxsep=0pt,left=3pt,right=3pt,top=3pt,bottom=3pt}

\newcommand{\sm}{Snowmass2021 }

\loadglsentries{tex/glossary}
\makeglossaries

\input{tex/workshopsymbols}

\newcommand\snowmass{\begin{center}\rule[-0.2in]{\linewidth}{0.01in}\\\rule{\linewidth}{0.01in}\\
\vskip 0.1in Submitted to the  Proceedings of the US Community Study\\
on the Future of Particle Physics (Snowmass 2021)\\
\rule{\linewidth}{0.01in}\\\rule[+0.2in]{\linewidth}{0.01in} \end{center}}

\begin{document}

%\pubblock

\title{Accessibility in High Energy Physics: Lessons from the Snowmass Process}

\author[1]{K.A.~Assamagan}%\thanks{ketevi@bnl.gov}}
\author[2]{C.~Bonifazi}
\author[3]{J.S.~Bonilla}%\thanks{johan.sebastian.bonilla@cern.ch}}
\author[4]{P.A.~Breur}%\thanks{sanderbreur@stanford.edu}}
\author[5]{M.-C.~Chen}
\author[6]{T.Y. Chen}
\author[7]{A.~Roepe-Gier}%\thanks{amber.roepe@ou.edu}}
\author[8]{Y.H.~Lin\thanks{\href{mailto:ylin@snolab.ca}{ylin@snolab.ca}}}
\author[9]{S.~Meehan}
\author[10,11,12]{M.E.~Monzani}
\author[13]{E.~Novitski}%\thanks{en37@uw.edu}}
\author[14]{G.~Stark}%\thanks{gstark@cern.ch}}
\affil[1]{Physics Department, Brookhaven National Laboratory, Upton NY, USA}
\affil[2]{ICAS-ICIFI-UNSAM/CONICET, Argentina and Universidade Federal do Rio de Janeiro, Brazil}
\affil[3, 8]{CERN, European Organization for Nuclear Research, Geneva, Switzerland}
\affil[4, 9]{SLAC National Accelerator Laboratory, Menlo Park CA, USA}
\affil[5]{Department of Physics and Astronomy, University of California, Irvine CA USA}
\affil[6]{Fu Foundation School of Engineering and Applied Science, Columbia University, New York NY, USA}
\affil[7]{ATLAS Experiment}
\affil[8]{Queen’s University, Department of Physics, Engineering Physics \& Astronomy, Kingston ON, Canada}
\affil[8]{SNOLAB, Creighton Mine \#9, 1039 Regional Road 24, Sudbury ON, Canada}
\affil[9]{2021-2022 AAAS Science \& Technology Policy Fellow}
\affil[11]{Kavli Institute for Particle Astrophysics and Cosmology, Stanford University, Stanford CA, USA}
\affil[12]{Vatican Observatory, Castel Gandolfo, V-00120, Vatican City State}
\affil[13]{Center for Experimental Nuclear Physics and Astrophysics, University of Washington}
\affil[14]{Santa Cruz Institute for Particle Physics, University of California, Santa Cruz CA, USA}

\renewcommand\Authfont{\fontsize{12}{14.4}\selectfont}
\renewcommand\Affilfont{\fontsize{9}{10.8}\itshape}

\maketitle

\vspace{-1cm}

\renewcommand*\abstractname{ABSTRACT}
\begin{abstract}
%\lead{CINDY}

\noindent Accessibility to participation in the high energy physics community can be impeded by many barriers.  These barriers must be acknowledged and addressed to make access more equitable in the future.  An accessibility survey, the Snowmass Summer Study attendance survey, and an improved accessibility survey were sent to the \sm community.  This paper will summarize and present the barriers that prevent people from participating in the \sm process, recommendations for the various barriers, and discussions of resources and funding needed to enact these recommendations, based on the results of all three surveys, along with community members' personal experiences.
\end{abstract}

%\textbf{Transfer stuff from \url{https://docs.google.com/document/d/1a_XqcI7r76Vo_Mjirxdzd_1fPoyHWqyV2m8wCdU38YA/edit##}}

\snowmass

%OUTLINE:

%What are the barriers HEP community members have to accessing Snowmass?
%By extension, participation in HEP community (e.g. conferences, meetings, etc.)

%What kind of improvements can be made to improve access?
%Are meeting transcripts sufficient?
%Is auto-captioning useful/sufficient?

%Resources and funding
%Who is responsible for providing access?
%Who is responsible for paying for access?
%Who is responsible for arranging for access?
%How to distribute information about the resources?

%Conclusion: Recommendations to HEP community planners and P5
%Checklists to consider for when planning meetings (hybrid, online, vs. in-person.  Duration of the event)
%Future work: linking demographics to accessibility to the field?

%For March 15th: mention that new results will be expected

\newpage
\glsunsetall
\begin{spacing}{0.1}
\tableofcontents
\end{spacing}
\glsresetall

\renewcommand{\thefootnote}{\arabic{footnote}}
\setcounter{footnote}{0}

\include{tex/introduction}
%\include{tex/barriers_in_hep}
\include{tex/improvements}

\include{tex/resources_funding}
\include{tex/logistics}
\include{tex/conclusion}
\include{tex/contributions}

%%%%%%%%%%%%%%%%%%%%%%%%%%%%%%%%%%%%%%%%%%

%  If you would like to use BibTEX for the bibliography, please feel free to do so.  It is not required.

%  To use BibTeX,

%    1.  uncomment the following two lines,
%    2.  comment out everything below from  \begin{thebibliography}{99}   to \end{thebibliography).
%    3.  create the file  myreferences.bib, and process this file in the usual way

\addcontentsline{toc}{section}{References}
\bibliographystyle{JHEP}
\bibliography{refs}  % file ref.bib

\clearpage
\printglossaries

%%%%%%%%%%%%%%%%%%%%%%%%%%%%%%%%%%%%%%%%%

\include{tex/appendix}

\end{document}

%% file: tex/glossary.tex
\newacronym{CERN}{CERN}{European Organization for Nuclear Research}
\newacronym{LHC}{LHC}{Large Hadron Collider}
\newglossaryentry{ATLAS}{name={ATLAS}, description={a general-purpose detector at the \gls{LHC}}}
\newacronym{HSF}{HSF}{\gls{HEP} Software Foundation}
\newacronym{IRIS-HEP}{IRIS-HEP}{Institute for Research and Innovation in Software for High Energy Physics}

\newacronym{DEI}{DEI}{Diversity, Equity, \& Inclusion}
\newacronym{HEP}{HEP}{High Energy Physics}
\newacronym{ADD}{ADD}{attention deficit disorder}
\newacronym{ADHD}{ADHD}{attention deficit hyperactivity disorder}
\newacronym{ADA}{ADA}{Americans with Disabilities Act}
\newacronym{SEC}{SEC}{Snowmass Early Career}
\newacronym{WER}{WER}{word error rate}
\newacronym{LHCP}{LHCP}{Large Hadron Collider Physics conference}
\newacronym{APS}{APS}{American Physical Society}
\newacronym{DPF}{DPF}{Division of Particles and Fields}
\newacronym{EAC}{EAC}{Ethics Advisory Committee}
\newacronym{CUWiP}{CUWiP}{Conference for Undergraduate Women in Physics}
\newacronym{NSF}{NSF}{National Science Foundation}
\newacronym{DOE}{DOE}{Department of Energy}
\newacronym{FASED}{FASED}{Facilitation Awards for Scientists and Engineers with Disabilities}
\newacronym{EF}{EF}{Energy Frontier of Snowmass}
\newacronym{WCC}{WCC}{White Coat Captioning}
\newacronym{d/D/HoH}{d/D/HoH}{d/Deaf/Hard-of-Hearing}
\newacronym{SL}{SL}{Sign Language}
\newacronym{ASL}{ASL}{American Sign Language}
\newacronym{PI}{PI}{Principal Investigator}
\newacronym{AI}{AI}{artificial intelligence}
\newacronym{API}{API}{Application programming interfaces}
\newacronym{APS-IDEA}{APS-IDEA}{American Physical Society Inclusion, Diversity, and Equity Alliance}
\newglossaryentry{steno}{name={steno}, description={short for stenographer, a person trained to type or write in shorthand methods, e.g. with a stenographer keyboard}}
\newacronym{CART}{CART}{Communication Access Realtime Translation}
\newacronym{RTSC}{RTSC}{Real-Time \Gls{steno}-Captioning}
\newacronym{STTR}{STTR}{Speech to Text Reporter}
\newacronym{ICHEP}{ICHEP}{The International Conference on High-Energy Physics}

\newglossaryentry{a11y}{name={a11y}, description={accessibility}}

%% file: tex/workshopsymbols.tex
\def\beq{\begin{equation}}
\def\eeq#1{\label{#1}\end{equation}}
\def\eeqn{\end{equation}}

\newenvironment{Eqnarray}%
   {\arraycolsep 0.14em\begin{eqnarray}}{\end{eqnarray}}
\def\beqa{\begin{Eqnarray}}
\def\eeqa#1{\label{#1}\end{Eqnarray}}
\def\eeqan{\end{Eqnarray}}

\let\bar=\overbar

\def\lsim{\mathrel{\raise.3ex\hbox{$<$\kern-.75em\lower1ex\hbox{$\sim$}}}}
\def\gsim{\mathrel{\raise.3ex\hbox{$>$\kern-.75em\lower1ex\hbox{$\sim$}}}}

\def\del{\partial}
\def\Dslash{\not{\hbox{\kern-4pt $D$}}}
\def\dslash{\not{\hbox{\kern-2pt $\del$}}}
\def\pslash{\not{\hbox{\kern-2pt $p$}}}
\def\ETmiss{\not{\hbox{\kern-4pt $E$}}_T}

\def\Dlr{\mathrel{\raise1.5ex\hbox{$\leftrightarrow$\kern-1em\lower1.5ex\hbox{$D$}}}}

\def\MSB{{\bar{M \kern -2pt S}}}
\def\msb{{\bar{\scriptsize M \kern -1pt S}}}

\def\drb{{\bar{\scriptsize D \kern -1pt R}}}

%% file: tex/introduction.tex
\section{Introduction}
\label{sec:introduction}
%\lead{CINDY}

\subsection{Overview}

\gls{DEI} are increasingly recognized as crucial issues in society at large, and physics\footnote{Physics is no exception.} lags behind in accessibility to people who are members of marginalized groups. The authors believe the physics community must actively protect people's fundamental right to participate in physics regardless of disability, identity, or background. %By creating an accessible and inclusive field, 
In addition, research shows that diversity works to make us smarter~\cite{HowDiversityWorks} and that socially diverse groups (i.e., diversity of race, ethnicity, gender, and sexual orientation) are more innovative than homogeneous groups. Through the process of understanding a variety of viewpoints and making the effort to reach a consensus, diverse groups are better at anticipating the needs required to achieve their goals.
  
In order to improve accessibility in the organizations that comprise the physics community, we must first understand what barriers people face and what resources are required to overcome these barriers. In this paper, a \textit{barrier} is anything that prevents or makes substantially more difficult to actively participate in physics activities. These barriers can include but are not limited to mental health, finance, time commitment, and physical constraints.

\begin{quote}
    \textbf{NB:} The terminology we use to describe different conditions in this paper are based on suggestions by community members and the \href{https://ncdj.org/style-guide/}{National Center for Disability and Journalism's Style Guide}. Wording choices also vary from country to country. We have tried to use preferred language but we might have made mistakes. Please feel free to reach out to us if we have used a word that is hurtful so that we can correct our error.
\end{quote}
  
\subsection{Scope of Paper}

This paper discusses the barriers that people have observed throughout the Snowmass process, the types of resources available, and recommendations for improving accessibility.

Data about barriers and suggestions for improvements were collected from multiple sources: three surveys, conversations with and written work by community members outside the context of the survey, the authors' experiences organizing accommodations, best-practices guidelines by physics and other organizations, and outside research. For more information on data sources, see~\Cref{sec:datacollection}.
The recommendations suggested here represent the consensus view of the authors arising from their analysis of information from all these sources; we do not claim endorsement by the respondents to the survey, other community members who provided input, or any other external organizations.

Though the surveys were of Snowmass participants about Snowmass activities, the recommendations for improving accessibility are made to apply to all \gls{HEP}. By recognizing and raising awareness of the various barriers and finding resources to mitigate these barriers, the authors aim to make these findings and recommendations a tool for making \gls{HEP} a more inclusive field and lead to impactful dialogue at the agency level and elsewhere.   

%There are many ways to promote this effort within any organization, from encouraging younger members to ask questions first after a presentation to ensuring equitable access to restrooms and quiet areas at a conference. This letter of interest is aimed to suggest a contributed paper that will document (perhaps in a checklist form) ways to improve and make Science more inclusive, with a light shown introspectively on the Snowmass 2021 process. This allows the contributed paper to be a formal part of the Snowmass 2021 process and hopefully lead to impactful dialogue at the agency level, and elsewhere.

%Dear colleagues,
%The Community Engagement Frontier's D&I group (led by Mu-Chun Chen, Samuel Meehan, Carla Bonifazi, Ketevi Assamagan) prepared a survey to collect inputs from the community on the issues of accessibility. This is primarily intended to be completed from the perspective of how accessibility applies to the Snowmass process for Snowmass21. It will provide valuable insights that can likely be extrapolated more broadly to our field but the inputs will most directly and immediately be used to affect the Snowmass process over the coming one and a half years.
%We would greatly appreciate it if you could take a moment to participate in this survey:
%https://docs.google.com/forms/d/e/1FAIpQLSfhe-kAC4xyqu9AsYyQn423kMlyjgaRsckEyDfazBIzbx3Xlg/viewform
%Best regards,
%Young-Kee Kim
%Chair, DPF

\subsection{Prevalence and Awareness of Accessibility Issues}

\begin{figure}[ht]
    \centering
    \includegraphics[width=\textwidth]{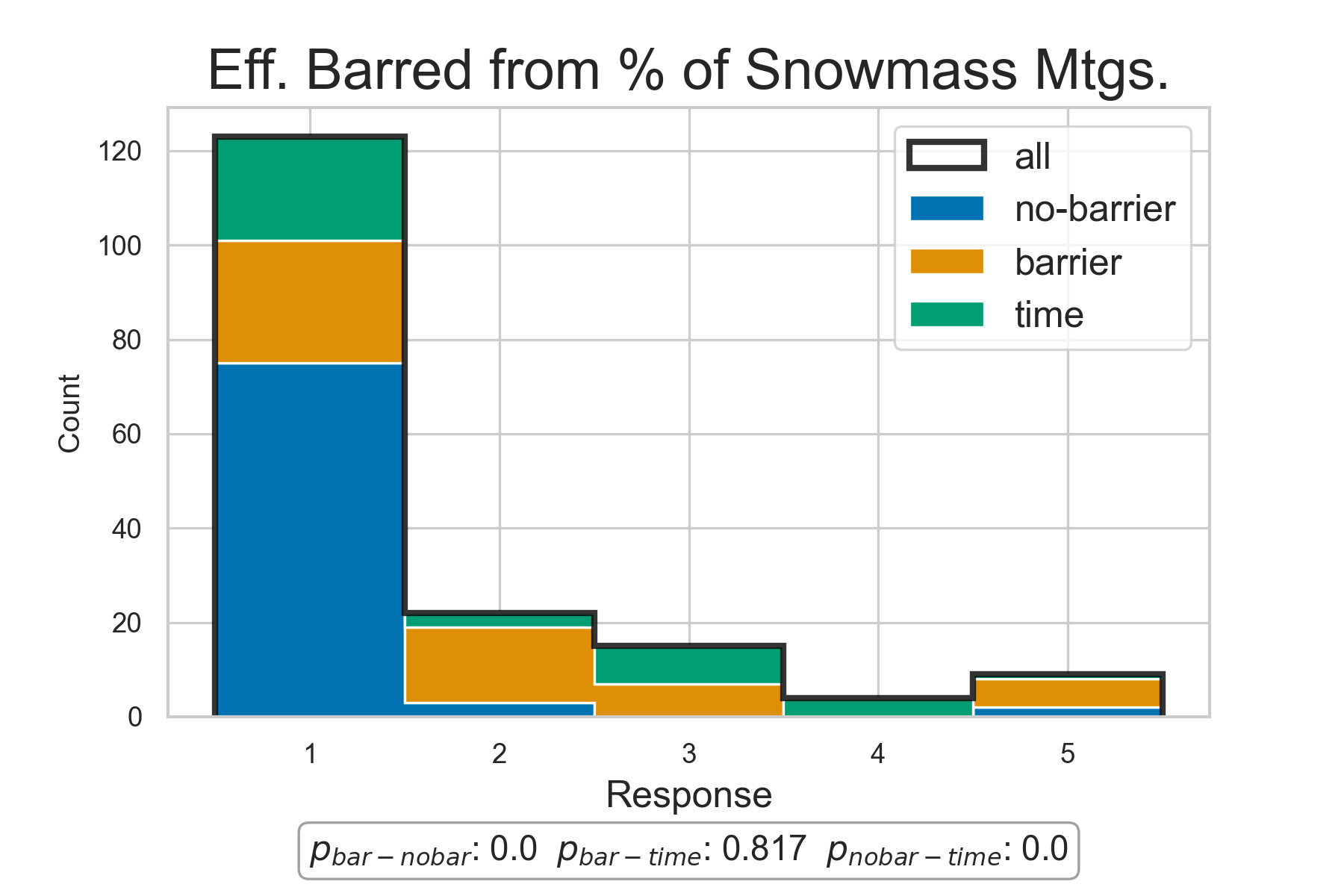}
    \caption{This bar graph shows the percentage of Snowmass meetings and/or activities have accessibility needs (if any) effectively barred the survey respondents from participation.  Response `1' is ``$<$20\%", response `2' is ``20-40\%", response `3' is ``40-60\%", response `4' is ``60-80\%", and response `5' is ``$>$80\%".  The $p$-values show that the response distributions between those who reported at least one barrier (bar) and those whose only barrier is time commitment (time) are not statistically significant.  However, the response distribution how the amount of meetings and activities impacted by accessibility needs are significantly different between population who reported no barriers (nobar) versus those who report at least one barrier, whether that barrier is solely time commitment or not.}
    \label{fig:eff_barr}
\end{figure}

About 20\% of those surveyed have personally experienced a barrier to participation in Snowmass. About 80\% of the people are aware of accessibility issues for others. The majority of respondents name lack of financial support, mental health issues, deaf/hard of hearing, and visual disability/blind as such problems. Over a quarter of the respondents name colorblindness and physical disability as barriers. About 20\% of respondents are not aware of any accessibility issues for others.  The impact of these barriers can be summarized in Figure~\ref{fig:eff_barr}, which reports in the percentage of Snowmass meetings and/or activities that the survey respondents have reported their accessibility needs (if any is reported) have barred them from participation.  The distribution of impact shows that those with barrier(s) are barred from participating in higher \% categories due to their barriers, whereas most respondents with no barriers see no impact on their participation in Snowmass.  

%Due to the pandemic, many people did not expect to have in-person meetings for Snowmass. About 15\% of the participants assume their accessibility issue will affect about 1 to 10 in-person meetings. Some of the accessibility issues, like physical or financial, can affect in-person meetings significantly more than virtual meetings. While, for example, audible and visual accessibility issues (or the lack of human interaction) are just as or more impactful on virtual meetings. For participants with an accessibility issue, over half of the respondents assume this will affect ten or more virtual meetings. About 5\% report that it will affect 50 or more virtual meetings. Several new questions were added to the second Snowmass accessibility survey to investigate the effectiveness of hybrid meetings.

The impact of these barriers extends beyond the people directly experiencing them. A second-order effect occurs with everyone's ability to collaborate with these individuals, which is the opposite of what we all want as a community. In this sense, the entire community profits from creating a more equitable environment for collaboration. In addition to directly benefiting individuals supported by improved access, we encourage everyone to focus on the community who profits from the interactions with that individual, which would otherwise be hindered or impossible were it not for this support.

\subsection{Doesn't the ADA Resolve These Issues?}

The \gls{ADA} of 1990 requires specific accommodations to “prohibit discrimination against individuals with disabilities in all areas of public life, including jobs, schools, transportation, and all public and private places that are open to the general public” ~\cite{ADA1991}. Its success in some respects, such as in the installation of wheelchair ramps in many new buildings, has led to an impression that the status quo of how things are done is in full compliance with this law. However, as one \gls{PI} who has considerable first-hand experience dealing with these issues highlighted in their response to the survey, this is not the case.

\begin{quote}
    “Every event ends up being different, which is quite a challenge. Even domestic events, which nominally should conform to ADA requirements, sometimes do not have (or are unwilling to acquire) funds to support necessary accommodations. Instead, these organizations require bailouts from the employing institution, larger organizations or funding agencies, or the supervisor's private research funds. International events are also consistent in being unwilling to provide funding for accommodations.”
\end{quote}

For US-based events, there are cases where the institution places an individual in the circumstance where they must choose to (1) provide support themselves or (2) bring their grievance to light in an official way. This situation may mean putting additional burden on a colleague or collaborator in the community. A cost-benefit analysis happens to determine if one must support the activities out of pocket or cause a disruption to an existing collegial setting. In the second case, this may mean no support is provided for the individual, so the person's right to accommodation is not fulfilled.  

So, although the \gls{ADA} is an outstanding achievement for the US, in our community it is not brought to bear in the way that one would hope it should be.

\subsection{Intersectionality of Barriers }

In addition to a singular accessibility need, it is \textit{essential} to note that one individual may have multiple needs. Just as those with multiple disenfranchised identities experience oppression that is more than just the sum of the parts, the same can be said about those with multiple disabilities or disabilities and other types of marginalized identities. Even if one type of need is met, it does not necessarily mean that everyone needing that accommodation will be automatically able to contribute fully. The types of accommodations that work best for each individual depend on their context and experiences.

%Of the respondents, one individual answered affirmatively to all of the choices on the ``What barriers exist for you?'' question, making it challenging to interpret because one answer was that they ``See no such barriers''. However, two participants answered with two needs, and in both cases, the second need pertained to mental health. In conjunction with the individual affirmative responses to this question about mental health, this response highlights that this is perhaps one of the more effective ways to impact the community.

%% file: tex/improvements.tex
\section{Barriers to Participation and Recommendations for Accommodations}
\label{sec:improvements}

Through the authors' experiences and feedback from our community, we have found that there are a number of hurdles that effectively bar members from being able to fully participate in the Snowmass process. This section will detail the barriers experienced by community members as well as recommendations on how to make a space inclusive for all members. These issues and recommendations are by no means exhaustive: they are based off of Snowmass community feedback collected through several surveys. The recommendations presented are also a \textit{starting point} for conference organizers and not a perfect solution for every individual. When the need for accommodation arises, it is crucial to confer with the individual in need to ensure the space is actually including them and that no financial burden of accommodation be placed on the individual. We also encourage event organizers to invite associated ethics groups to review the plans to ensure that the responsibility is distributed amongst conference leadership, thereby minimizing oversights. \textbf{No one knows a person's needs better than themselves, and that should be respected at all times.}

\subsection{Financial}

A considerable focus of Snowmass is interaction and collaboration at in-person events. Although COVID-19 is modifying how we collaborate and reducing the number of events that are able to be in person, financial barriers often create two communities - the ``haves,'' and the ``have nots''. These hurdles can be related to the size of an individual's home institution and the available funds in their grant. A lack of financial stability can also disproportionately affect \gls{SEC} members, who will ultimately help execute the ``30-year vision''. However, they may not be able to participate in these events because, in the words of survey respondents, the \gls{SEC} members ``do not talk about it'' or because ``it is not worth them going''.

\subsubsection{Survey Results}
A summary of the second Accessibility Survey can be found in Figure~\ref{fig:eff_barr} and Appendix~\Cref{app:updated_survey_results}. 

In the first accessibility survey (~\Cref{sec:firstaccessibilitysurvey}), 5\% of people reported that their participation in Snowmass was personally affected by lack of financial support, and 50\% reported that this was a barrier they were aware of others experiencing.

The Snowmass 2022 Summer Attendance Survey (\cite{SnowmassSummerStudyAttendanceSurvey2022}, ~\Cref{sec:attendancesurvey}), respondents were asked whether funding would affect their participation in the Summer Study. 11\% said funding concerns would likely prevent them from attending, 26\% said it might prevent them from attending, 16\% said they could attend but it might limit their participation, and 45\% said it was not a concern. Related suggestions from commenters included: provide a remote or hybrid attendance option, make plenaries available online, make it possible to participate for partial meeting, provide childcare assistance (financial and logistical) [see ~\Cref{sec:caretaking}], provide financial assistance with conference costs, provide refunds if COVID-19 prevents travel, and support participants required to isolate because of COVID-19.

\subsubsection{Recommendations}

Our recommendation is to provide limited travel grants advertised by conference organizers made available for individuals through an application procedure overseen by an ethics group associated with the conference (be it the \gls{DPF}, \gls{EAC} or an institution-based group). These grants could be similar to student travel awards advertised by \gls{APS} and could be available to scientists at all stages of their careers. One form of travel grant that has been used successfully, for the US-ATLAS Computing Bootcamp in August 2019 and other events in our community, is to award amounts based on matching the estimated cost of travel for an event. This ensures that the individual traveling and their home institution have buy-in on the individual's participation in the event.

Conferences should also strongly consider making their entrance fees sliding-scale or waived for under-resourced and early-career scientists.

\subsection{Caretaking Responsibilities}
\label{sec:caretaking}

Though Snowmass relies on the interactions that are facilitated by in-person meetings, they can present challenges for those who are caregivers for others, such as children or dependents with special needs. This disproportionately affects researchers with young children, who are themselves disproportionately early-career researchers who most are most affected by the decades-long vision that is the product of Snowmass.
Therefore, it is particularly important to offer assistance so that researchers with dependents can participate on an equal footing in Snowmass.

\subsubsection{Survey Results}

The Snowmass 2022 Summer Attendance Survey (\cite{SnowmassSummerStudyAttendanceSurvey2022},~\Cref{sec:attendancesurvey}), did not directly ask about the impact of caretaking responsibilities on participation in the Summer Study, but it did ask what factors might prevent or limit participation, with ``Competing responsibilities'' as an option, which includes caregiving. 15\% said competing responsibilities would likely prevent them from attending, 27\% said they might prevent them from attending, 30\% said they could attend but it might limit their participation, and 29\% said it was not a concern. Comments suggested that organizers provide childcare assistance (financial and logistical), avoid weekends, make it possible to participate for a partial meeting to accommodate schedule constraints, provide a remote or hybrid attendance option, and make plenaries available online.

\subsubsection{Recommendations}

Conferences can accommodate caregiving responsibilities by providing childcare or by supporting the travel of an accompanying person.
In both situations, some extra funding should be budgeted by the conference to cover, at least partially, those costs.
A conference can also provide a private, clean room for breastfeeding or pumping.

\subsection{Mental Health}

Pre-existing mental health issues can affect a person's experiences in physics, just as they can in every other field and profession. These mental health issues can be exacerbated by the field's climate, particularly in the context of pressure to participate in a high volume of meetings and the anxiety that can accompany feeling ``fear of missing out.'' These issues, particularly the latter, are troubling throughout the Snowmass process, which only increases the necessary bandwidth required for people to meaningfully contribute
Both community members and survey respondents reported that these issues have affected their ability to participate in Snowmass and contribute to shaping the field for the next thirty years. Early-career scientists report that they perceive themselves as being in a doubly-precarious situation, juggling responsibility for carrying out research and driving the field forward with uncertainty about employment prospects and difficult power dynamics.

These issues might not be externally visible, but they impede physicists' work and negatively affect their personal lives \cite{Beretz}. Scientific progress should not come at the expense of an individual’s well-being - this is the framework from which we approach radiation safety and with which we should approach mental health.

\subsubsection{Survey Results}

In the first accessibility survey (~\Cref{sec:firstaccessibilitysurvey}), 8\% of people reported that their participation in Snowmass was personally affected by mental health, and 55\% reported that this was a barrier they were aware of others experiencing.
Respondents reported depression, anxiety, and \gls{ADHD}\footnote {ADHD is considered a disability in the United States and is described in the DSM-5 Diagnostic and Statistical Manual of Mental Disorders, 5th edition.} among mental health issues they personally experience.

In the Snowmass 2022 Summer Attendance Survey (\cite{SnowmassSummerStudyAttendanceSurvey2022}, ~\Cref{sec:attendancesurvey}), 16\% of respondents reported that they would be helped by a ``Quiet Zone'' near the meeting rooms where people can go to rest without interaction.

\subsubsection{Recommendations}

The recommendations here come in the form of cultural and leadership paradigm shifts that should be recognized:
\begin{itemize}
\item{At in-person events, the organizers should create a quiet space where people can withdraw to find peace and solace. These isolated areas are akin to the ``coffee spaces'' that exist but different from that in that they would be reserved for no speaking or interaction. These areas would allow the individuals not to venture far from the scientific discourse to recharge before continuing with the science.}
\item{Meeting chairs and event hosts should reflect on whether a meeting is ``actually necessary.'' Creating additional meetings does not always increase productivity, and based on this survey can often serve just to increase the required bandwidth, which can grind our community down. In this sense, being part of the solution is as simple as not being part of the problem.}
\item{Meeting participants should actively follow the code of conduct and strive to create a welcoming intellectual space.  Recognize that depression ``just makes everything hard,'' which may negatively impact others in invisible ways.}
\end{itemize}

\subsection{Physical}

Physical and mobility limitations can affect participation in in-person meetings. Within the US, the \gls{ADA} provides a framework that organizers of in-person meetings and their institutions are bound to follow. However, \gls{ADA} compliance both is often incomplete and can fail to fully meet the needs of all people with disabilities.

There are many types of physical disabilities beyond those that involve using a wheelchair. We must take into account the needs of those including, but not limited to,

\begin{itemize}
    \item People who use wheelchairs
    \item People who are limited in their walking speed or distance
    \item People for whom standing for long periods is not possible
    \item People who, due to chronic pain, require different seating than standard furniture
    \item People who have short or tall stature
    \item People who have upper-extremity disabilities, making it difficult to lift or carry objects
    \item People who require service animals
    \item People who require dedicated space for self-administering medication
    \item People who have strict dietary restrictions
\end{itemize}

\subsubsection{Survey Results}

In the Snowmass 2022 Summer Attendance Survey (\cite{SnowmassSummerStudyAttendanceSurvey2022}, ~\Cref{sec:attendancesurvey}), 9\% of respondents said they would benefit from there being seating available at all events, breaks, meals, discussion areas. 3\% reported that they would need assistance in traveling around campus, and 1\% said they would need specific furniture. Related comments included: plan breaks that are long and frequent enough, ensure mobility-related accessibility of buildings and accommodations, hold all events close to each other and to accommodations, have good chairs as well as some tables and standing desks in meeting rooms, provide private space for medical procedures, accommodate dietary restrictions, and plan frequent and long-enough breaks.

\subsubsection{Recommendations}

Our goal should be to ensure that people with physical disabilities are truly accommodated, with guaranteed, low-friction, dignified access to all aspects of the conference experience. The following statements by one person with mobility limitations illustrate the costs of not doing so.
\begin{itemize}
    \item ``Most receptions are ``stand-around and mingle.'' This is just a no-no for wheelchair folks (because of lack of conversation at eye level) and other limited mobility folks (we need to find a chair at some point and end up spending the evening alone in a corner)."
    \item ``Yes, we can hire a bus to ferry people around, but if I have to call someone asking to schedule a ride for me at a certain hour, there is extra work that I have to do to attend a session. In other words, the barrier for being driven around should be comparable to the barrier for just walking over there."
    \item ``People tend to be gracious if I ask them to skip in line at lunch or to hand me something I can’t reach. But the whole point of accessibility is that it’s a human right, not a favor that people “grant me” out of the goodness of their heart."
    \item ``People for whom fatigue makes extended meetings without breaks impossible.'' 
\end{itemize}

The following guidelines for accommodations are drawn from suggestions by physicists with physical disabilities, most of which were provided outside the context of the survey.\footnote{In addition to input from the authors, conversations with community members, and the surveys, these guidelines also draw from the work of A. Peet, e.g., \cite{PeetOnDisability}.}
\begin{itemize}
    \item For check-in, coffee, and meals
    \begin{itemize}
         \item It is good to have lots of stations rather than a small number of long lines.
         \item Having seating in the waiting area for at least some of these stations is helpful for people who cannot stand for a long time.
         \item Having a designated accessible line can be helpful for some people. Others might prefer not to use such a line, so it is best to make all the lines as accessible as possible.
    \end{itemize}
    \item Moving around within a building
    \begin{itemize}
        \item Organizers should arrange meeting locations (even in nominally \gls{ADA}-compliant buildings) to minimize the following:
        \begin{itemize}
            \item  Building entrances, hallways, internal doorways, and electric door-opening buttons can be obstructed by furniture or construction work or broken.
            \item A single floor of a building can be split-level, requiring steps to get from one part to another.
            \item Meeting rooms with stadium seating might not have both levels accessible to wheelchairs and might present an impediment to asking questions of speakers after a talk.
            \item Accessible paths between conference activity sites and between these sites and restrooms, both within and between buildings, are sometimes much longer than non-accessible paths, making it difficult to transition between activities in the time allotted.
            \item Items can be out of reach due to height for those in wheelchairs or with short stature (such as coat hangers in bathrooms, coffee at the coffee break, food, water fountains.)
        \end{itemize}
         \item Conference organizers can reduce these barriers by surveying all paths between conference areas several times and correcting any problems they see. The following schedule of surveys is recommended:
         \begin{itemize}
            \item  During the room-booking and agenda-setting process, to ensure that accessible paths exist, to make maps of these paths to post to the website and to distribute to attendees, and to make estimates of the transit times needed to get between events
            \item A couple of weeks before the conference begins to allow time to arrange for any needed repair work to be done
            \item A day or two before the conference starts to make sure no obstructing objects have appeared and to put up additional signage along paths
        \end{itemize}
    \end{itemize}
    \item For getting between buildings
    \begin{itemize}
        \item As much as possible, having conference events at a single building is good.
        \item Provide temporary disabled parking at conference building(s) for those in need.
        \item It is best to have a bus always there ready to transport people between events; the next best is to have a convenient app for summoning a bus; the worst is to step out of the session to call for a bus on the phone.
        \item The bus needs to be a kneeling bus.
        \item List distances between buildings on the conference website.
    \end{itemize}
    \item Hotels
    \begin{itemize}
        \item Make sure there are rooms available with beds accessible to people with short stature and people in wheelchairs (e.g., not most dorm beds).
        \item If possible, have hotel rooms available within walking distance\footnote {What the correct "walking" distance is depends on the accommodation need of the specific person in mind.} of the main conference site. The next best option is to have shuttles available on-call, and the next best option after that is to have them at scheduled times.
    \end{itemize}
    \item Eating
    \begin{itemize}
        \item Food should be catered (brought to the building) or available in an in-house food court, not requiring walking to restaurants.
        \item Buffet-style food should be served from multiple stations (see above), or food should be served at tables.
        \item At least some seating should be available at all breaks and social events.
        \item At the reception/dinner, have things arranged and spaced well enough that people in wheelchairs or with mobility restrictions can move around to at the very least their table, the restroom, and places to get food. Having tables arranged in a ring around the edge of the room can work well to enable access more broadly throughout the room.
    \end{itemize}
    \item Meeting Rooms
    \begin{itemize}
        \item Arrange seating within rooms such that there are places for people to put their bags and coats where they do not block the aisles.
        \item Make sure there is room for wheelchairs to pass, and to sit once in the room, as well as for service animals.
        \item For some people, having a desk or spot at a table is helpful. Even having a modest amount of seating this way in each room, reserved for those who need it most, would help.
        \item Conference organizers should survey participants ahead of time about their needs for modified furniture and organize the provision of this furniture.
        \item Set up a designated chat-with-the-speaker-after-their-talk area with seating just outside each meeting room, and have it be the norm that discussion happens seated in that area. Properly-designed spaces allow all people (including those who need to sit, are in wheelchairs, or have short stature) to see and be seen and fully participate in these post-talk discussions.
    \end{itemize}
    \item Designated organizer/volunteer availability
    \begin{itemize}
        \item To the extent possible, accommodations should be arranged ahead of time and made easily available to attendees without volunteer help. However, for complications that come up, an easily identifiable person or people should be available to troubleshoot with attendees.
    \end{itemize}
\end{itemize}

%Universities are not used to accommodating temporary visitors. For example, university athletic departments have access to equipment, such as golf carts. The host institution can at least make the facilities they provide their athletic teams available at nominal cost. \FIXME{remove or integrate above}

\subsection{Auditory}
\label{subsubsec:auditory}

Due to the international nature of the physics community (which was exacerbated by the COVID-19 pandemic), most physics meetings worldwide have migrated to taking place online.
Being \gls{d/D/HoH}\footnote {For essential distinctions between deaf vs. Deaf, we encourage the reader to review Chua, Smith, et al.~\cite{ASEEPeer32676} to understand the socio-linguistic and cultural differences between the two groups.} can drastically affect participation in both in-person and virtual events. Moreover, because there is no broad education within the community about these challenges, a dichotomy of perceived and real challenges has emerged. This frequently leads to issues not being realized or addressed until very late in the organization of an event, leading to a solution that is unsatisfactory or simply exclusionary.
For those without auditory barriers, microphone quality is still a problem in some cases where it makes it very hard to follow the speaker.
However, a much bigger question is: ``How can these meetings be accessible for people with auditory barriers?''

As with other accommodations described in this paper, each person is the best expert on their own needs, and it is best to ask rather than to make assumptions. For example, some \gls{d/D/HoH} people can read lips and therefore appear to interact with non-\gls{d/D/HoH} people effectively when in person. However, due to poor video quality, they cannot transfer this skill to the virtual world. Therefore, the intuition of ``but I can talk to that person in real life easily enough'' should not be carried over to the virtual world. 

Similarly, there are a variety of ways of providing auditory-related support, e.g., auto-captioning, human-provided live captioning, human-provided post-pacto transcriptioning, \gls{ASL} interpretation, assistive listening devices, etc. Which technique(s) are appropriate for a given event is crucially dependent on the particular context. An example: not all \gls{d/D/HoH} people know \gls{ASL}, so it is important to not assume that providing an \gls{ASL} interpreter will meet everyone's needs. The following sections provide information to help event organizers plan appropriately.

\subsubsection{Capabilities of Different Auditory Accommodations}

\textbf{\gls{ASL} vs. Live Captioning}

\gls{ASL} is a language unto itself that uses different structures and vocabulary from the spoken language, such as English, in which captions are provided - they are two different languages. So \gls{ASL} interpreters cannot be viewed as ``replacements'' for live captioning, and vice-versa; they serve different roles. 

Captioning is good for making presentations accessible and for capturing single-channel discussions where all questions are audible to the captioner. In practice, this means that it works well for making talks more accessible and for capturing virtual group discussions. Captioning can be done remotely even for an in-person meeting, as long as the session chair ensures that all questions are asked with mics or are repeated so as to be audible to the captioners. However, captioning does not cover informal conversations at in-person conferences. For some \gls{d/D/HoH} participants, in-person ASL interpretation might be appropriate to fulfill that need.

\textbf{Live Captioning vs. Transcriptions}

Event organizers sometimes ask, \textsl{Can I meet this need by providing a transcript for the meeting?} While transcripts make the recording of the meeting accessible after the fact, which is excellent, they do not encourage or enable participation \textbf{during} a meeting. 
Allowing everyone to participate in a meeting actively requires additional resources such as real-time \gls{steno}-captioning or \gls{ASL}.

\textbf{Auto-C(r)aptioning Is Not Accurate Enough}

Another common question is, \textsl{Is auto-captioning by a computer/\gls{AI} (affectionately referred to as ``auto-craptioning'') sufficient?} The quick answer is No\footnote{Auto-captioning is as useful as the LHC is for measuring the triple Higgs coupling.}. Some of the many problems with auto-captioning are the inability to recognize specialized jargon, training bias towards English-speaking white males, and not picking up on sarcasm or emotions, all of which have led to the impossibility of being part of a conversation~\cite{AudioAccessibility, TheAtlantic, DCMP, ReelWords, A11yNYC, Wired, AngryDeafPeople}.

One of the essential quantitative metrics to sufficiently evaluate live captioning and transcriptions is \gls{WER}, which measures the mistranslation rate of spoken words. As a benchmark, a \gls{WER} of 15\% roughly translates to an error for every 2.4 seconds of speech. Alternatively, if one reads a book, the equivalent is three words wrong and/or missing per sentence. For auto-captioning/AI-based captioning, this assumes ``white, American, tech CEO voice'' and worsens significantly if the person speaking has an accent or the conversation includes technical jargon, both of which are prevalent in our community. The rev.com service provides the industry-leading autocraption service with a \gls{WER} of 16.6\%. Other services include Google (\gls{WER}=18\%) and Otter.AI (\gls{WER}=20\%), which have begun to be leveraged by some of the frontier groups for their working group meetings. To get a sense of the difference in these services as compared to a human transcription, which is sub-percent \gls{WER}, we encourage you to compare the captions found in a YouTube recording portion of an internal \gls{ATLAS} SUSY summary meeting~\cite{YouTubeSUSY2020}, which has both Google auto-captions by selecting ``Subtitles → English (auto-generated)” versus the \gls{WCC} subtitles which can be selected with “Subtitles → English (United States)'', also documented in~\Cref{app:autocraption_susy}. This can be augmented by viewing the subtitles on the Otter.AI generated transcript for an Energy Frontier working group meeting from June, also documented in~\Cref{app:autocraption_ef}. Turn the sound off, watch the videos with captions, and ask whether crucial points are missed due to the poor \gls{WER}. Rev.com does not do live autocraptioning but rather auto transcription, providing only the text after the event. Google does provide live autocraption, but it depends on the platform. Otter.AI is the only one that has tried to provide more opportunities to use its service via an \gls{API} connection.

\subsubsection{Costs}

\textbf{ASL interpretation: \$75-\$200/hr.} ASL interpreters typically charges \$75-\$100/hr per hour per person, and they often work in teams of two. Keep in mind that any travel and hotel costs must also be covered.

\textbf{Human professional live captioning: \$80-\$250/hr.} Captioners who focus primarily on tech jargon (e.g., \gls{WCC}) can achieve and sustain a \gls{WER} around 0.01\%-0.5\% consistently, even with accents. Their cost is at the higher end of this price range, usually around \$2-4/min.

\textbf{Human professional post-facto transcription services (transcript becomes available after a delay of hours or days): \$45-\$180/hr.} rev.com is \$1.25/minute and is currently the service of choice for the \gls{HSF} Training group and \gls{IRIS-HEP} organization to caption training videos as this training on using the Docker application. They do a pretty good job, and their \gls{WER} with human transcriptioners is around 4-5\%. 3PlayMedia and Scribie are around the same \gls{WER} of 4-5\% and cost \$3/min and \$0.75/min, respectively. 

\textbf{Live or post-facto autoc(r)aptioning: \$3-9/hr.} rev.com costs \$0.15/minute, the Google service's cost is \$0.048/min, and Otter.AI costs around \$0.06/min. 

\textbf{Volunteer post-facto transcription by physicists: equivalent of over \$200/hr.}
One approach that has been to lower costs is to use autocaptioning, then to have those captions corrected by volunteers from within the \gls{HEP} community who are not professional captioners. In this case, the cost comes from their volunteered time. Roughly speaking, it takes a factor of 8-10 to manually correct a transcript or caption file for normal humans (every minute is 8 minutes of volunteer human work to fix); see the Case Studies in the following section for examples supporting this factor. If an autocraptioned presentation is later fixed by humans, the cost is at least \$150, not explicitly including the sunk cost due to volunteers' time to fix the transcripts. If we include labor costs for the volunteers, this will add an additional cost of roughly \$65/hr of each volunteer's time.\footnote{Labor costs were determined as follows: Average salary for a full professor of physics at a 4-year public institution according to the \href{https://www.aip.org/statistics/salary-calculator}{American Institute of Physics's Salary Calculator} is \$134,200. Dividing this by a total of 2080 work hours in a calendar year (40hrs/week for 52 weeks) gives an hourly rate of \$64.52.} There is also additional overhead in making these captioned videos available. In the context of Snowmass, this is also the time that ``should'' be spent focusing on investigating and developing our scientific priorities! If one hires a professional\footnote{Alternatively known as a \gls{steno}-captioner.} outright, such as \gls{WCC}, this will cost approximately \$40-80 per 20 minutes.

\subsection{Case Studies}

In this section, we give examples of several conferences' experiences providing auditory accommodations. The authors wish to recognize the efforts of the organizers and volunteers at these conferences, and to stress that they do not aim to single out these particular conferences for criticism for shortcomings in coverage. Accessibility problems at \gls{HEP} meetings are pervasive, and these issues are common to many conferences beyond those mentioned here; by including these examples, we hope to help the community understand the challenges faced and benefit from others' experiences. The community as a whole must increase its financial and logistical support for accommodations to improve on this situation, and this includes improving support for organizers of conferences to enable them to provide better accommodations.

\emph{Case Study: \gls{LHCP}}

\begin{quote}
    In 2020, when the pandemic initially broke out in North America, \gls{LHCP} 2020 was the first major \gls{HEP} conference that was organized after the start of pandemic, with free registration. This meant the budget and the time to pivot the conference format from in-person to virtual were both limited. The \gls{LHCP} organizers found funding for and provided 16.5 hours of professional live-captioning.  These transcripts were available in the conference Indico agenda after the conference, but were not able to be included in the recordings due to technical reasons.  All talks (82.5~hours of content) were recorded, autocaptioned, and available after the conference.  The organizers arranged for volunteers to correct autocaptioned presentations for 30-40 additional talks (10~hours), with dual goals of increasing accessibility to those particular talks and providing a training data set for improving real-time captioning AI. On Twitter, two volunteers documented their experience correcting autocaptions (\url{https://twitter.com/freyablekman/status/1273354943026679808}). Each of these two volunteers spent 3 hours of their time on correction for each 20-minute presentation.  These corrections were included in the recordings available on the CERN Document Server and linked from the Indico agenda.  72.5 remaining hours of recorded talks are available online with only uncorrected autocaptioning.
\end{quote} 

\emph{Case Study: \gls{SEC}-Inreach}

\begin{quote}
    The \gls{SEC}-Inreach group of \sm organized a colloquium series as one aspect of their professional development activities for early-career Snowmass participants. The first edition of this series results in 1h20m of content, which was auto-captioned. \gls{SEC}-Inreach received feedback after the event that the \gls{WER} of these captions were too high for participants to be able to follow the discussion as it happened live. \gls{SEC}-Inreach then uploaded a recording of this colloquium to YouTube, and a team member revised the auto-captions. This effort took approximately 6-8 hours of focused work performed by a physicist rather than a certified or trained professional transcriber. The \gls{SEC}-Inreach group is actively searching for any resources possible to hire live \gls{steno}-captioning services for the remaining events in this series. Their efforts to procure funds have not yet been successful.
\end{quote}

\emph{Case Study: \gls{ICHEP}}

\begin{quote}
    The \gls{ICHEP} conference made the entirety of their plenary and public events live-captioned by \gls{WCC}.
\end{quote}

\emph{Case Study: 2022 Snowmass Summer Study}

\begin{quote}
    The Snowmass Summer Study is providing live captioning by \gls{WCC} for all plenaries and two tracks of parallel sessions for the 10-day conference. It is also providing in-person ASL interpreters to all participants who request it as an ADA accommodation. During presentation-intensive sessions, one ASL interpreter per requester is booked to cover informal conversations between captioned talks. During social activities, two ASL interpreters per requester are booked to enable sustained discussion. The lead ASL interpreter is someone with extensive physics interpreting experience who traveled from out of town, and the second interpreter is a highly experienced local interpreter without physics-specific experience. The ASL interpreters were organized by a liaison at the host institution's Disability Services Office.
    
    Obtaining funding (10s of k\$) for these accommodations was challenging. While the \gls{NSF} has a way to provide accessibility funding (See Sec. \label{sec:funding})), this mechanism is not well-known, and by the time they became aware of it, organizers had been awarded all the NSF funding for the Snowmass Summer Study earmarked solely in the form of participant support, which cannot be used for this purpose. The host institution did not provide funding because in-person participation in the Summer Study is only for paying attendees rather than being free and open to the public. The organizing committee eventually secured funding for these services as part of the main \gls{DOE} contribution to the Snowmass Summer Study and through a dedicated grant from a private foundation.
    
    Booking service providers far enough ahead of time\textemdash and assessing the scale of the need early enough to make accurate bookings\textemdash was a significant challenge. Because it was such a large commitment for them, \gls{WCC} asked for confirmation and a binding contact several months before the start of the summer study, before registration had opened. The reservations therefore had to be made on the basis of the anonymous attendance survey.
\end{quote}

\subsubsection{Survey Results}

% \FIXME{In the first accessibility survey (~\Cref{sec:firstaccessibilitysurvey}), X\% of people reported that their participation in Snowmass was personally affected by being \gls{d/D/HoH}, and Y\% reported that this was a barrier they were aware of others experiencing.}

In the Snowmass 2022 Summer Attendance Survey (\cite{SnowmassSummerStudyAttendanceSurvey2022}, ~\Cref{sec:attendancesurvey}), 14\% of respondents indicated that at least one of the hearing-related accommodations would be helpful. 2\% indicated that live \gls{steno}-captioning or \gls{ASL} interpretation would be necessary for their full participation, and 1\% said an assistive listening device or amplification service would be helpful.

\subsubsection{Recommendations}

Currently, the \sm \gls{DEI} group recommends announcing virtual meetings early enough to allow time for arranging accommodations in conjunction with any individual with these needs. In effect, the individual must provide the accommodation their institution currently arranges. For in-person meetings, the \textbf{assumption} is that the hosting institution will bear the costs and provide all necessary accommodations under the \gls{ADA}. However, this is not always the case. The survey revealed that the institution that employs these individuals must bear the cost. In both cases, the community burdens the individual and the institution. However, this view is in contrast with what the community feels \textbf{should} be the case and given that these services benefit \textbf{everyone} in the community not only through the intellectual collaboration with those who have the need but secondary benefits:

\begin{enumerate}
\item The ability to view events without sound or perform
\item A text-based search of an event to find specific sections
\item No need for explicit proceedings -- as having good \gls{steno}-captions makes it trivial to generate proceedings for an event automatically
\end{enumerate}

As these services benefit the entire community, the entire community should bear the costs in a distributed way.
A detailed cost analysis was performed (~\Cref{sec:closed_captioning_cost}).
We propose these costs come from a collective budget arranged by the \gls{DPF} Executive Committee. These expenses need to be taken into account for the next Snowmass proceedings and similarly for other \gls{HEP} conferences and planning activities.
Over the next 30 years, we will spend multiple billions of dollars on scientific research. As such, the community needs to ensure that a fraction of that amount gets set aside for equitable planning, which is the bare minimum our community should commit to doing.

Furthermore, we propose that a mechanism exists for an individual or frontier convener to request the service to be arranged for all future Snowmass proceedings, which facilitates the arrangement of a live \gls{steno}-captioner by the convener responsible for holding that meeting. The organizers will publish this mechanism on the Snowmass site, and Frontier Conveners will be authorized to make this request utilizing these funds. The request mechanism needs to be public and easy to find to minimize the burden we place on an individual. Such requests should also be commonplace when people are registering for a conference or other community event.

\subsubsection{Recommendations for Organizers}

There are a number of matters that have arisen during the Snowmass 2021 proceedings about those who are \gls{d/D/HoH} that we feel would help anyone organizing conferences.

\textbf{Starting early.}
The tasks for planning audio accommodations, in approximate sequential order, include: figuring out who will attend, determining which services would meet those needs, finding providers, getting cost estimates, securing funding, booking services, and working with attendees and service providers to ensure that final preparation and the conference go well. The time needed for each step scales with the scale of the conference in length, complexity, and size, and requires months for a Snowmass-Summer-Study scale conference.

\textbf{Deciding what form of accessibility support to provide.} 
In nearly all cases, guidance should be from the person requesting as they will tell you their preferred communications~\cite{chua2017behind}. Not all \gls{d/D/HoH} people know American Sign Language (\gls{ASL}) because they may not have had access to this growing up or are fluent in another signed language such as LSF (langue des signes française; French Sign Language). Some d/HoH folks do not sign and are oral-only communicators; some other Deaf folks only communicate via a signed language (e.g., \gls{ASL}), and others are a hybrid of both. One can make two approaches here: \textsl{reactive} (wait for an accessibility request) and \textsl{proactive} (provide access no matter what).

In the case of being proactive, providing \gls{steno}-captioning is a solid choice. While \gls{ASL} (or a signed language) is more inclusive for Deaf signers, captioning will help more than just \gls{d/D/HoH} and is a little easier to arrange or budget. Captioning would help anyone who cannot hear well (age, accent, environmental noise, other factors), has trouble constantly focusing/listening, or those for whom English is not their first language, and finds it easier to follow if there is written English. This is a utilitarian approach.

Conferences should be prepared to provide both \gls{steno}-captioning and \gls{ASL} interpretation. Note that there are a lot of contextual clues at play here when comparing \gls{ASL} and \gls{steno}-captioning - enough that it is essentially apples to oranges. For example, California-based interpreters do better with Asian-centric accents because they have grown up or have gotten accustomed to those accents. Usually, there is a regional bias in play for that sort of thing where an \gls{ASL} interpreter is often in person from the same area where the request happens -- so they overcome that initial barrier of regional dialects -- while an online/remote \gls{steno}-captioner could come from anywhere.

\textbf{Finding service providers.} The most skilled providers might have schedules that fill up ahead of time, might charge more, or might (in the case of in-person work) need to travel from another city. Getting captioners and ASL interpreters who are familiar with physics content is quite rare but can provide much better service. A few suggestions follow for finding service providers.

The person making the request for the services might have recommendation of providers they've worked with before or heard about.

The Disability Services Office or similar office at the host institution might have local networks to tap into. 

Networks of providers, such as \gls{CART}/\gls{RTSC}/\gls{STTR}, can all vary widely (the pool is a lot smaller than \gls{SL} interpreters), and the quality depends on the provider's background knowledge, audio quality, and any accents.
        
In the United States there are a few leading agencies to go for when requesting \gls{steno}-captioning. If it is a Science-Tech-Engineering-Arts-Math (STEAM) jargon-heavy meeting or conference, we recommend \href{https://whitecoatcaptioning.com/}{White Coat Captioning} - also suitable for international. For general purpose meetings where it is not very jargon-heavy or technical, we recommend \href{https://www.interpreter-now.com/}{Interpreter Now} or the host institution's staff captioners\footnote{If the institution has any, but not very typical.}. If it is very, very last minute, typically anywhere from 30-minutes notice to 24 hours notice, we recommend \href{https://www.acscaptions.com/}{ACS Captions} or \href{https://www.captionfirst.com/}{CaptionFirst} as they are most likely to find someone that fast. 

Internationally, both \gls{WCC} and \href{https://www.ai-media.tv/}{AI-Media} are best. However, other geopolitical factors might come into play such as \href{https://gdpr-info.eu/}{GDPR}, privacy concerns, international conflicts, or other cultural issues.

For sign language interpreters -- internationally, there is only one agency we recommend: \href{http://www.overseasinterpreting.com/}{Overseas Interpreting} (\gls{ASL}, and a few other signed languages supported) for in-person interpretation. When we are talking about VRI (Video-Relay Interpreting), such as for remote/online, the only agency we recommend there with any semblance of consistency in the quality is \href{https://www.interpreter-now.com/}{Interpreter Now}.

\textbf{Incorporating service providers' work requirements.} Captioners require safety breaks (e.\~g.\~, 10 min per hour or 15 min per 90 min) and might need to work in teams to trade off if continuous captioning is needed or for particularly long shifts (>4 hours). ASL interpreters typically work in pairs in some situations, such as when continuous interpretation is needed for more than about an hour or when the interpreter is speaking for the participant (doing ASL-to-spoken-English translation) in addition to doing spoken-English-to-ASL interpretation. Sometimes captioners and interpreters charge by the minute or hour for short chunks of time (1-2 hours), but for longer shifts it might be necessary to book a full day, especially to account for the service provider's prep and setup time.

\textbf{Enabling with providers' preparation.} Captioners are able to enter specialized terms into their software and improve their accuracy if they have access to slides, notes, or other materials ahead of time. ASL interpreters can also prepare by learning specialized signs for the subject material. Service providers need to be compensated for this time and organizers should work with presenters and people requesting accommodations to facilitate this prep work.

\subsection{Visual}

The reliance on computers in physics can present challenges for those with visual disabilities, in particular for those who are blind, have low vision, or are colorblind. This section discusses aspects of presentations and web design that can be barriers. There are additional accommodations at in-person conferences that can be useful for people who are blind or have low vision. The authors have not yet gotten specific feedback from the physics community about these needs, but we include general resources on them in the recommendations section.

\subsubsection{Survey Results}

% \FIXME{In the first accessibility survey (~\Cref{sec:firstaccessibilitysurvey}), X\% of people reported that their participation in Snowmass was personally affected by being visually impaired or blind, and Y\% reported that this was a barrier they were aware of others experiencing. Z\% of people reported that their participation in Snowmass was personally affected by colorblind, and ZZ\% reported that this was a barrier they were aware of others experiencing.}

In the Snowmass 2022 Summer Attendance Survey (\cite{SnowmassSummerStudyAttendanceSurvey2022}, ~\Cref{sec:attendancesurvey}), 4\% said they would be helped by the use of colorblindness-friendly color schemes in presentations, and 1\% by alt-text describing images in presentations. Write-in comments included requests for organizers to require green laser pointers and provide guidance on how to use colorblindness-friendly color schemes.

\subsubsection{Recommendations}

There are many resources on the internet that already provide suggestions for accommodating the blind/low vision community. The American Foundation for the Blind has posted a host of actions people can take in order to make online meetings more accessible, including the article \href{https://www.afb.org/blog/entry/accessibility-online-conferences-classes}{5 Accessibility Actions You Can Take When You’re Moving Your Conference or Classes Online}. There is also a list of suggestions from the Perkins School for the Blind on in-person accessibility: \href{https://www.perkins.org/resource/make-your-meeting-accessible/}{Make Your Meeting Accessible}.

Physicists who are blind or have low vision often use tools to make computers accessible to them, such as non-visual desktop access, speech-based job parsing, and screen reader tools that allow them to hear the contents rather than see them. The physics community should adopt norms that improve accessibility for those using these tools, e.g., by consistently including descriptive alt-text for images in documents. Moreover, for LaTeX, one of the more common tools in our field and particularly so for theorists, such text → auditory auto-translations fail\textemdash this is an issue. The University of Nevada, Reno has outlined a way to translate LaTeX equations to HTML5 to ensure a screen reader can interpret it properly.\footnote{Making LaTeX Math Equations More Accessible.\\ \href{https://www.unr.edu/accessibility/resources/documents/accessible/research-and-publications/strategy-1}{https://www.unr.edu/accessibility/resources/documents/accessible/research-and-publications/strategy-1}} This is a primary example of a small step that can be taken to ensure our science is readable for all.

Another common visual disability is colorblindness, which affects 1 in 12 men and 1 in 200 women worldwide. This can make it difficult for individuals to interpret plots that use unfavorable color palettes with low contrast or do not leverage the use of other stylistic traits to differentiate between results on a single plot. This can impede understanding may dissuade a person from engaging in the discussion if they cannot correctly interpret a plot during a talk. Addressing this issue in a sweeping way is challenging due to the many types of colorblindness that exist, but online tools (e.g., We Are Colorblind ~\cite{wearecolorblind}, Colblinder ~\cite{colblinder}) can mimic colorblindness to help in creating readable plots. There are already some \gls{HEP} collaborations that incorporate colorblind needs into their style guides \cite{belleII2021}. The \href{https://stash.desy.de/projects/B2D/repos/belle2style/browse}{Belle-II collaboration plotting style guide} explicitly states that plots and graphics must be ``color-blind friendly'' and provides example palettes for analyzers to use. Our recommendation for the community is to provide a set of guidelines and tools, such as the existing Belle-II color palette, to conference organizers and participants in an easy-to-use fashion for conference materials and presentations, and that event organizers communicate the expectation that these be used universally.

%% file: tex/resources_funding.tex
\section{Resources and Funding}
\label{sec:funding_resources}

The authors argue that, even outside the US and the scope of the \gls{ADA}, the responsibility falls with the organizers of a meeting, conference, or workshop to ask who needs access support and then provide that support.

\subsection{Data from the First  Accessibility Survey}

We asked these final sets of questions to understand where the responsibility lies to find a solution for an accessibility need, financial or otherwise. This section contained questions with a series of four statements to ascertain how much the respondent agreed with where their funding comes from, or should come from:
\begin{itemize}
    \item The individual with the accessibility need
    \item The home institution of the individual
    \item A professional organization (i.e. \gls{APS} or a subsidiary like \gls{DPF}) 
    \item (For international events) The host institute/country
\end{itemize}

The respondents showed that just under 70\% rely on either their home institutes or their personal/group grants to provide access. About 10\% rely on a third-party such as a professional organization or the host institute, and about 20\% tell us they rely solely on themselves to provide accessibility needs. When asked who should be responsible for providing access, most argue that \gls{APS} should be responsible for the individual's needs. Almost everyone agrees it should not be the responsibility of the individual.

\subsection{Who is Legally Responsible for Paying for Access?}

\textbf{Financial burden of providing access should never fall on the individual requesting it.}

As several of the authors have encountered in attempting to arrange for disability accommodations at Snowmass and other events, there is widespread confusion in the community about when accommodations are legally required under Section 504~\cite{504} and the \gls{ADA} at US conferences, what constitutes sufficient accommodations, and who bears the legal responsibility for funding them. Some basic answers from \cite{reasonableaccommodations}\footnote{Note: this source also contains advice for relying on amateur captioning and interpretation and for burdening \gls{d/D/HoH} conference participants with planning their own accommodations. The authors strongly disagree with these suggestions, for reasons explained in~\Cref{subsubsec:auditory}.}:

\begin{quotation}
    \noindent ``What is a public accommodation?''  This is important since according to The Americans with Disabilities Act (1990) (\gls{ADA}: \cite{ADA1991}):  Entities that are considered ``public accommodations'' must provide reasonable accommodations for individuals with disabilities.
    ``Within U.S. law, public accommodations are generally defined as facilities, both public and private, that are used by the public. Examples include retail stores, rental establishments and service establishments, as well as educational institutions, recreational facilities and service centers''~\cite{wikipediaPublicAccommodations}.  For more information, see \cite{EEOCADAQA}.\\

    \noindent The next terminology to understand is the term ``private club'' since private clubs and religious entities are not considered public accommodations.  If your club allows members of the public to join, doesn't limit its total membership, is not overly stringent regarding membership requirements, and doesn't require applicants to be personally recommended, sponsored by or voted on by current members then it is likely that your organization is not going to be considered a private club and will therefore be considered a ``public accommodation'' \cite{privateClubsLaborLaw}\\
    
    \noindent According to U.S. Department of Justice guidelines (\gls{ADA}: \cite{ADA1991}) if your organization is open to the public, even if you have ``qualifications" for membership,'' then it is likely that yes your organization is a public accommodation.\\
    
    \noindent [...]\\
    
    \noindent For Deaf and hard of hearing people ``reasonable accommodations'' are typically considered to consist of an interpreter and/or real-time captioning.\\
    
    \noindent [...]\\
    
    \noindent As far as reasonable accommodations provided by organizations hosting conferences -- whether or not an accommodation is (legally) considered reasonable will vary depending on the size and budget of the ``organization'' (which can be thought of as the ``tenant'') and to some degree the the owner of the facilities (the ``landlord'').
\end{quotation}

Similarly, according to conversations with one US university's disability services office, the organization that puts on the conference is typically responsible for ensuring access to participants with disabilities. If a host institution is a co-sponsor\textemdash for example, if it collects, manages, and/or profits from the registration fees, or if it provides space without charging rental fees\textemdash it might also bear partial responsibility.

For Snowmass, note that while people organizing the public/open meetings are responsible for getting the resources in place, the fiscal cost falls on \gls{APS} and \gls{DPF}\footnote{Note that as \gls{DPF} is a part of \gls{APS}, it is ultimately governed by \gls{APS} constitution/bylaws - it cannot operate as an independent organization.}. This should be Snowmass global policy that each topic and frontier has point people — because this is not just a matter of being decent but also following U.S. law. In particular, if we look at the organization and how Snowmass fits in, Snowmass is affiliated with \gls{DPF}, which is under \gls{APS}. \gls{APS} is the parent organization when it comes to Snowmass. \gls{APS} receives both \gls{NSF} and \gls{DOE} funding (citation needed). \gls{APS} is legally required to cover fiscal costs because while the 1990 American with Disabilities Act (ADA)~\cite{ADA1991} is relevant here, Section 504~\cite{504}, which predates ADA, is relevant, via ``\textsl{any program or activity receiving federal financial assistance or under any program or activity conducted by any Executive agency or by the USPS}'' (citation needed). 
As the \gls{APS} receives federal funding for a subset of its activities, such as the \gls{CUWiP}, it is covered by Section 504 to provide accessibility for all of its activities. As \gls{DPF} falls under \gls{APS}, \gls{DPF} is also covered by Section 504.

\subsection{Sources of Funding for Accommodations}
\label{sec:funding}

The \gls{NSF} offers \gls{FASED}~\cite{NSFFASEDinstructions, NSFFASEDannouncement} awards, described as follows:
\begin{quotation}
    \noindent As part of its effort to promote full utilization of highly qualified scientists, mathematicians, and engineers, and to develop scientific and technical talent, the Foundation has the following goals:
    \begin{itemize}
        \item to reduce or remove barriers to participation in research and training by persons with physical disabilities by providing special equipment and assistance under awards made by \gls{NSF}; and
        \item to encourage persons with disabilities to pursue careers in science and engineering by stimulating the development and demonstration of special equipment that facilitates their work performance.
    \end{itemize}
    
    \noindent Persons with disabilities eligible for facilitation awards include \gls{PI}s, other senior personnel, and graduate and undergraduate students. The cognizant \gls{NSF} Program Officer will make decisions regarding what constitutes appropriate support on a case-by-case basis. The specific nature, purpose, and need for equipment or assistance should be described in sufficient detail in the proposal to permit evaluation by knowledgeable reviewers.\\
    
    \noindent There is no separate program for funding of special equipment or assistance. Requests are made in conjunction with regular competitive proposals, or as a supplemental funding request to an existing \gls{NSF} award.
\end{quotation}

These proposals can be submitted via \href{https://research.gov}{research.gov} by selecting \gls{FASED} as the proposal type \cite{newProposalType}.

\subsection {Example Cost Calculation for Closed Captioning}
\label{sec:closed_captioning_cost}
Ballpark calculations needed to assess expected costs\footnote{Regardless of cost, there is a matter of legal rights, which necessitates urgent application to \gls{NSF} for funds.} for captioning all Snowmass meetings will quickly occur to any physicist who has spent some time musing about the various energy scales by which the laws of our Universe are applicable. Suppose each current frontier/topical group defined in the Snowmass 2021 process has biweekly meetings that need to be covered. In that case, that will be an expected 3000 meetings throughout the current iteration of the Snowmass process. Assuming each meeting is roughly an hour, assuming \$10 per minute for captioning the meeting brings us to an expected \$1.8m cost in providing captioning services. However, this entire scenario obscures or hides the additional benefit-cost: the benefit of the meeting vs. the cost of people's time to attend the meetings. If one sees value in having biweekly meetings, great, but if many of those biweekly meetings for a particular topic or frontier often end short or without contributions, then perhaps one needs to rethink the number of additional costs for arranging that meeting. Accessibility requests are just a way of making many of those hidden costs more obvious.
  
Lastly, there are additional benefits to things like captioning that one would get for free, such as a post-meeting transcript\footnote{This also means that note-taking is less of a chore, and perhaps can automatically be done; with ML? Less work!}, the ability for many people to attend the meeting while muted and still follow along, or if they are in noisy situations, for students who have difficulty focusing, for people who need to step out for a second but still be able to catch up, etc. Furthermore, there is also the excellent benefit of just providing thousands of pages of transcripts directly to the funding agencies at the end of the Snowmass process with every discussion that happened -- documented in writing -- verbatim. Then one can take tools to categorize, sort, and maybe start quickly gleaning common concepts or ideas that spring out of multiple meetings that do not overlap enough, etc. The sky is the limit, and the entire Snowmass process becomes transparent. Suddenly, that \$1.8m cost does not seem bad for reducing a significant amount of labor for everyone. If we made this effort entirely proactive, there is no need for:
\begin{itemize}
    \item a \gls{d/D/HoH} person to request captioning,
    \item the organizers to ensure it is in place for a specific meeting, and
    \item the point-of-contact \gls{a11y} person to handle fiscal costs continuously, by wrapping it into a single contract that establishes the status quo for the entire Snowmass process.
\end{itemize}

%% file: tex/logistics.tex
\section{Logistical Recommendations for Conference Organizers}
\label{sec:logistics}

\subsection{Bare Minimum Best Practices}

\subsubsection{Budget for Accommodations}
In order to make physics truly accessible, it's necessary to have a shift in mindset such that accessibility isn't regarded as an option to be included if the budget allows it, but a base necessity, just like a room for people to sit in and a projector for the slides. If you didn't have the budget for those things, you would consider yourself to not have the budget to hold your event. Similarly, if you don't have the budget to accommodate the needs of the attendees, then you don't have the budget for your event.

\subsubsection{Plan Ahead} 
Some accommodations require significant advance notice to book services or equipment. Organizers should plan far enough ahead to assess participants' needs, then reserve the services early enough. Longer events can require more notice. For example, for captioning, an approximate rule of thumb is that a 2-hour meeting should have at least a week's notice, and a 40-hour (full-week conference) should have four months' notice.

\subsubsection{Have a Single Point of Contact}
A good design pattern is to designate a single person as a point of contact for all things accessibility (\gls{a11y}); or a few people depending on the size of the event. A point of contact is someone involved in organizing the event to talk to for accessibility-related logistics. This person helps reduce much labor on everyone's part. The organizer does not need to learn how to arrange access for individuals. The requester does not need to explain or teach each organizer what access they need or what details to use. That way, instead of having to one-off to different people, one can communicate with the same person and say, ``this one too.'' The labor involved is much lower if the requester knows they do not have to find an organizer and convince them that this is their job, but instead email whoever is the designated person. A single point-of-contact is an all-around win-win solution that is rare in a process like Snowmass, which deals with many trade-offs. This has already been successful for conferences organized by some of the authors.

\subsubsection{Educate Yourself about Etiquette}
Consult resources like \cite{disabilityEtiquette} when training volunteers to learn more about how to be helpful to people with accommodation needs in an effective and respectful way. If someone with a disability corrects you on terminology or their needs, do not take it personally. Acknowledge and thank them for the correction and remember for the future.

\subsubsection{Solicit and Respect the Input of the People who Need Accommodations} Each individual knows their own needs best.

\subsection{Case Study: US ATLAS Annual Meeting Checklist}

As part of any meeting or event, in particular, one should strive to maintain a consistent set of ''live" materials that evolve and improve with experience and usage. A primary goal of this whitepaper should be to produce a checklist that can be provided to all event organizers to go through and ensure equitable access to meetings that promote inclusivity and foster a diverse environment. Below, a case study is provided from the U.S. ATLAS organization, which has started to provide a checklist to be used for all of its annual meetings. This checklist removes the mental strain of trying to remember all the details as a swamped, over-worked organizer, and the goal is to make it as easy as possible to make the event as inclusive as possible. The checklist was beta-tested during the 2019 US ATLAS annual meeting~\cite{USATLASAnnualMeeting2019}. This checklist is reproduced in~\Cref{app:checklist}.

%% file: tex/conclusion.tex
\section{Recommendations and Future Work}
\label{sec:conclusion}

%\lead{Cindy}

A survey was conducted within the Snowmass community in 2020 to find out the barriers in the context of accessing Snowmass2021 (and by extrapolation, the \gls{HEP} community), what existing resources are needed to overcome these barriers, and the sources of support required.  80\% of survey respondents reported that they see accessibility issues for others, which says that there is a need for resources and support to make Snowmass (and by extension, \gls{HEP}) more accessible for all.  Some of the barriers that have been reported include (but not limited to) financial, caretaker responsibilities, mental health, physical, and virtual access. Moreover, any individual might experience multiple barriers and might be affected differently by those barriers based on their experiences and identities. This means that making physics accessible requires addressing all barriers holistically and listening to the people who need accommodations about what works best for them

Currently most of the respondents rely on home institutes or their
personal/group grants to provide access; however, most argue that \gls{APS} should be responsible for providing support and that it should not be the responsibility of the individual.  

Based on our finding, we suggest that conference organizers go through the US \gls{ATLAS} Annual Meeting Checklist (or similar lists where applicable) well before a potential event to ensure that accessibility needs are met.  This includes planning ahead, surveying needs when surveying interest in the conference and/or meeting, requesting funding to support the needs based on the feedback, and booking the services to address the needs.  

While our surveys have shown the need for significantly improved accommodations, their limitations prevent them from answering certain important questions. Because survey respondents are not guaranteed to be representative of the \gls{HEP} community, the surveys don't reveal the overall prevalence of accessibility issues. Future studies with anonymous demographic information questions will not only make it possible to assess whether the surveyed population is representative, but will also be able to measure the ways in which accessibility issues might disproportionately affect different demographic groups.  

%% file: tex/contributions.tex
\section{Data Collection and Acknowledgements}
\label{sec:contributions}

\subsection{Data Collection}
\label{sec:datacollection}

In addition to drawing from best-practices guidelines from external organizations and other external research\textemdash which are cited in-line as they appear in the text\textemdash information were collected for this paper by survey and by experience of the authors, as outlined below.

\textbf{First accessibility survey (complete)}
\label{sec:firstaccessibilitysurvey}
% June 23rd, 2020
In June 2020, the Community Engagement Frontier: Diversity and Inclusivity subtopical group surveyed the Snowmass community to understand how accessibility impacted participation in the Snowmass21 process.  The digital survey was primarily distributed on the Snowmass Slack workspace and sent on several listservs. A total of 157 people filled in the complete survey, which represents 7\% of the 2244 members participating on Snowmass2021 Slack workspaces of August 1st. % The survey aimed to determine what accessibility people need for the Snowmass process, especially during the pandemic where more and more virtual meetings occurred. 
The survey consisted of 14 questions, of which some were open-ended and some were multiple-choice.  This questionnaire is in~\Cref{app:original_survey}.  While the responses provide an idea of the barriers in our field, they are by no means fully encompassing of all of the accessibility needs.  
%The primary assumption under which conclusions are drawn and extrapolated is that the respondents represent the individuals who have an accessibility need. 
The data, which includes names and emails collected from the original survey, are only available to the authors of this paper and the survey review committee.  The identifying information is optional, only used for further clarification.  
%Participants were informed that the survey was primarily intended to be completed from the perspective of how accessibility applies to the Snowmass process for Snowmass21 and aimed to provide valuable. %insights that can likely be extrapolated more broadly to the field of fundamental physics. 
Only aggregated data are presented in this paper. For responses and description of experiences included here, identifiable details were removed. 

\textbf{Attendance survey}
\label{sec:attendancesurvey}

The Snowmass Summer Study Local Organizing Committee (LOC) conducted a survey \cite{SnowmassSummerStudyAttendanceSurvey2022} in February 2022, with two goals: to assess how many people were likely to attend the summer study, and to quantitatively understand accessibility needs in order to start booking services and equipment. This survey was done independently of the Community Engagement Frontier: Diversity and Inclusivity subtopical group, though one author of this whitepaper (EN) is also on the LOC and was involved with this survey. A predecessor document to this whitepaper, including results of the first accessibility survey, was an important source for the creators of the attendance survey to understand what kind of accommodations should appear as options.

\textbf{Second accessibility survey (ongoing)}
\label{sec:secondaccessibilitysurvey}

Since more people have begun participating in Snowmass2021 since 2020, Community Engagement Frontier: Diversity and Inclusivity subtopical group is conducting an additional survey. It is similar to the first survey, with modifications to increase the ease of filling out the survey and analyzing responses, was sent out in March 2022.  The goal of the new survey is to gather more easily quantifiable responses, address questions that arose from the first survey, and increase the number of respondents to a more representative percentage of the Snowmass2021 population.  We will update this paper with the results from the new survey in July 2022.
The raw data will not be seen by anyone except for ARG, who has had training on research ethics, which included some human subjects research ethics.  The required questions will be aggregated into compiled data and presented in the paper.  The additional text responses will only be included if the survey respondents give explicit permission and serve as a way for the community to voice their concern/feedback.  We are not collecting any additional personally identifying information so the individual responses will remain anonymous.

\textbf{The authors' experiences organizing accommodations}

The authors' experiences have informed this paper. Some of the authors have organized accommodations for Snowmass events, \gls{CUWiP}s and a satellite conference aimed at people with disabilities and other marginalized identities. Several have experience working on accessibility issues while serving on ethics committees, the DPF Executive Committee, and on the Diversity, Equity, and Inclusion committees of their collaborations and departments, including one in \gls{APS-IDEA}. Other experiences include work on an accessibility working group and leadership of a student union. Multiple authors also have personal experiences using accommodations.

\subsection{Acknowledgements}

The authors would like to thank Josephine Paton for her input on the review of language usage, Daria Wang for her support and coordination through the contribute paper process, and the \gls{APS} \gls{DPF} and SLAC FPD for providing financial support for captioning at contributed paper meetings. All authors acknowledge conversations with community members who have personal experience with accessibility barriers, which provided substantial input for this document.

%% file: tex/appendix.tex
\begin{appendices}
    \crefalias{chapter}{appchap}
    \input{tex/app_checklist}\clearpage
    \input{tex/app_original_survey}\clearpage
    \input{tex/app_updated_survey}\clearpage  
    \input{tex/app_plots}\clearpage  
    \input{tex/app_autocraption_comparison_susy}\clearpage
    \input{tex/app_autocraption_comparison_ef}
\end{appendices}

%% file: tex/app_checklist.tex
\section{US ATLAS Annual Meeting Checklist}
\label{app:checklist}

The checklist below is copied verbatim from~\cite{USATLASChecklist} as a guide for all hosts of the US ATLAS annual meeting every year to follow to ensure equitable, diverse, and inclusive meetings. This checklist summarizes the steps to be taken by the organizing committee to ensure that we maintain consistent accessibility for all US ATLAS-sponsored conferences. In addition, US ATLAS \gls{DEI} contacts sent these guidelines to the ATLAS \gls{DEI} contacts as (hopeful) a stepping point for ensuring consistent accessibility for international conferences as well.

  \begin{enumerate}
    \item Organizing Committee
    \begin{enumerate}[label=\alph*.]
      \item Announce policy for selection of organizing committee (with call for volunteers)
      \item Represent the diversity of US ATLAS on committee (career stage, institution type, gender, race, ethnicity, disability, etc.)
      \item Ensure that at least one graduate student and one postdoc are included
    \end{enumerate}
    \item Registration Form
    \begin{enumerate}[label=\alph*.]
      \item Collect demographic information for annual review
      \item Include comment box for accessibility requests
      \item Agreement to Code of Conduct as part of registration process
    \end{enumerate}
    \item Website
    \begin{enumerate}[label=\alph*.]
      \item Include Code of Conduct~\cite{CERNCoC}
      \item Include all policies (instructions for speakers, instructions for session chairs, scholarship and award criteria)
      \item Include Equity and Inclusion strategies in speaker and chair instructions
      \item Provide information for local childcare options
      \item Provide point of contact for accessibility requests
      \item Provide organizing committee contact information with the associated areas of responsibility per person in case questions or concerns arise
    \end{enumerate}
    \item Venue
    \begin{enumerate}[label=\alph*.]
      \item Ensure accessibility of meeting rooms and bathrooms (e.g. for wheelchairs)
      \item Ensure availability of gender-neutral bathrooms
      \item Ensure flexible seating/free-standing chairs in meeting rooms (e.g. to accommodate wheelchairs) and that the spaces are scent-free
      \item Ensure that area for speaker can accommodate different heights and abilities (check podium height)
      \item Ensure that audio equipment is available and functions correctly
    \end{enumerate}
    \item Resources
    \begin{enumerate}[label=\alph*.]
      \item Offer student grant travel scholarships
      \item Offer childcare and room for breastfeeding if requested
      \item Set aside funding for accessibility requests
      \item Secure funds for outside speaker for Diversity, Equity and Inclusion talk
      \item Provide preferred pronoun stickers for nametags
    \end{enumerate}
    \item Program
    \begin{enumerate}[label=\alph*.]
      \item Devote 1-2 hours for advancement of understanding of equity and inclusion
      \item Provide professional development opportunities like career planning for students and mini-workshops on giving presentations, interviewing, grant-writing, application-writing or management style (etc.)
      \item Leave ample time for discussion and respect the timeline
      \item Have people state their name when asking a question or making a comment
    \end{enumerate}
    \item Closeout
    \begin{enumerate}[label=\alph*.]
      \item Report to IB with meeting demographics and fulfillment of D\&I guidelines
    \end{enumerate}
  \end{enumerate}
  
  

%% file: tex/app_original_survey.tex
\section{Accessibility of Snowmass Survey - June 2020}
    \label{app:original_survey}
    The following five pages contain the first accessibility survey (complete) that was sent out to the Snowmass community to understand how accessibility impacted participation in the Snowmass21 process. 
    
    \includepdf[pages={1-5},scale=0.8,pagecommand={}]{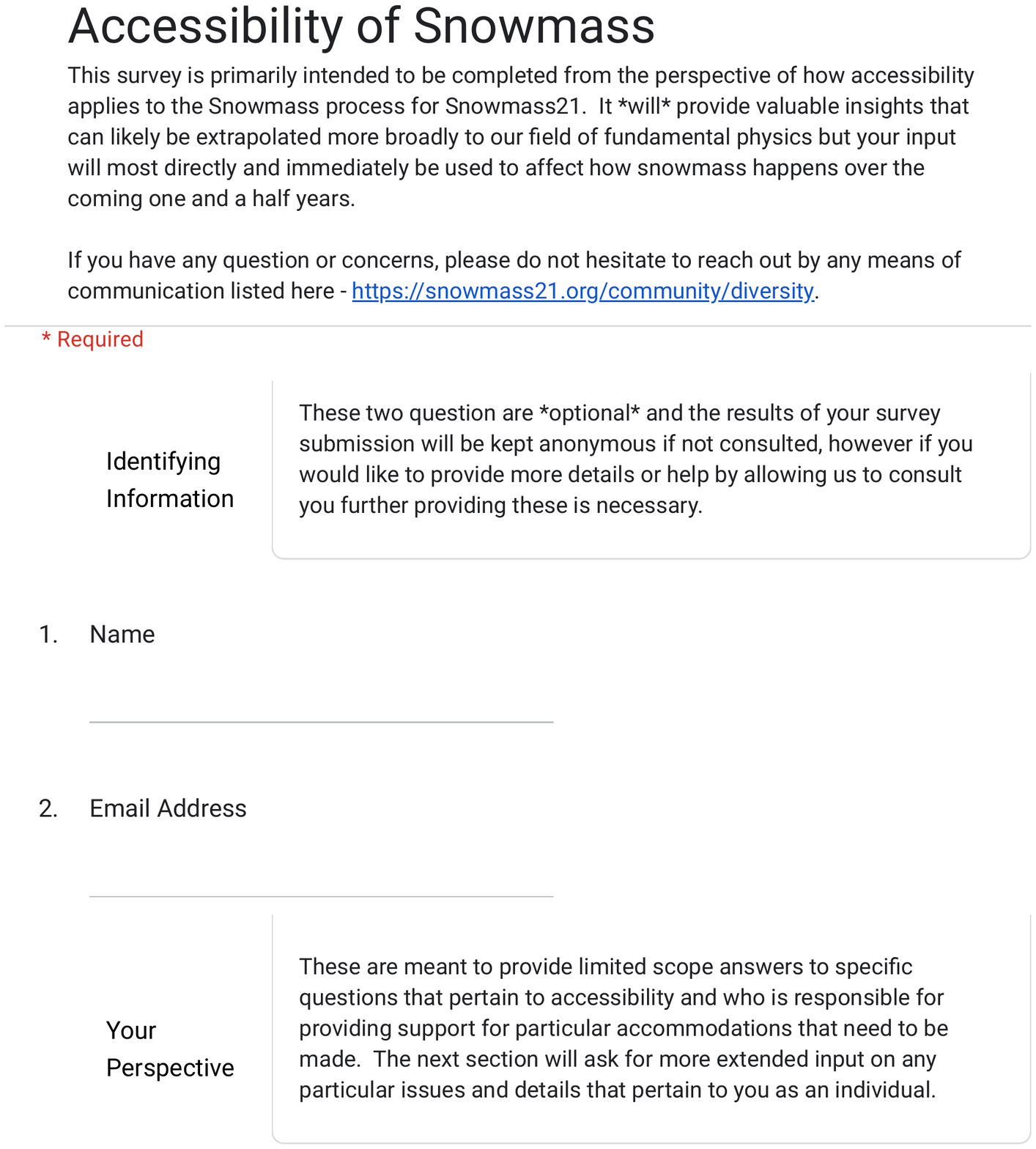}

%% file: tex/app_updated_survey.tex
\section{Accessibility of Snowmass Survey - March 2022}
    \label{app:updated_survey}
    The following five pages contain the second accessibility survey (complete) that was sent out to the Snowmass community to understand how accessibility impacted participation in the Snowmass21 process.  
    
    \includepdf[pages={1-7},scale=0.8,pagecommand={}]{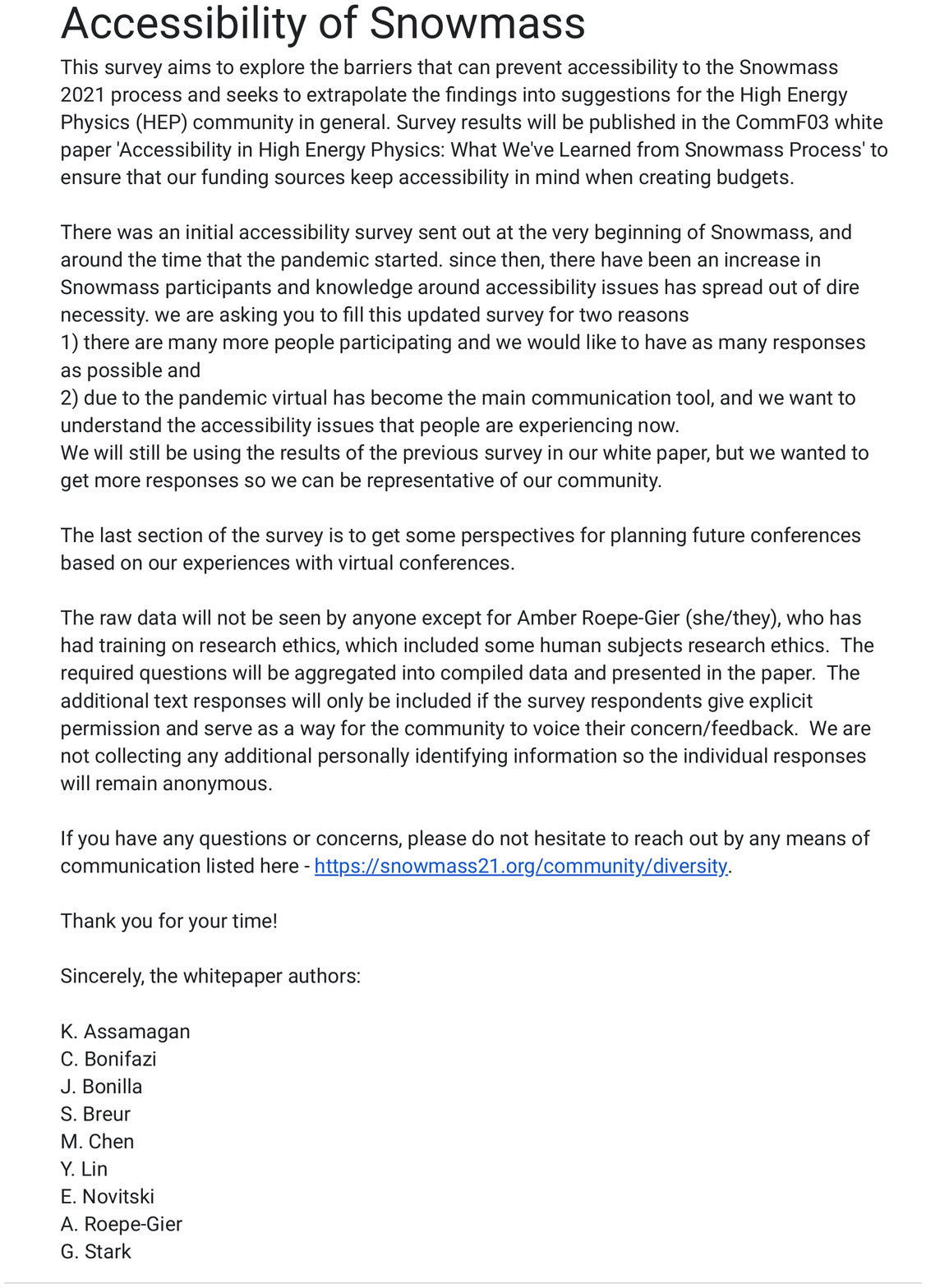}

%% file: tex/app_plots.tex
\section{Results from Accessibility of Snowmass Survey - March 2022}
    \label{app:updated_survey_results}
    This section contains the results from the second accessibility survey~\ref{app:updated_survey}.  173 respondents participated in the survey.  In the following section, `nobar' refers to respondent population that reported no barrier, `time' refers to respondent population that reported time commitment as their only barrier, and `bar' refers to respondent population that reported at least 1 barrier (not including respondents in `time' group).  The results reported in this section do not include the text answers provided by the respondents; those are included in the main text.

\begin{figure}[ht]
    \centering
    \includegraphics[width=\textwidth]{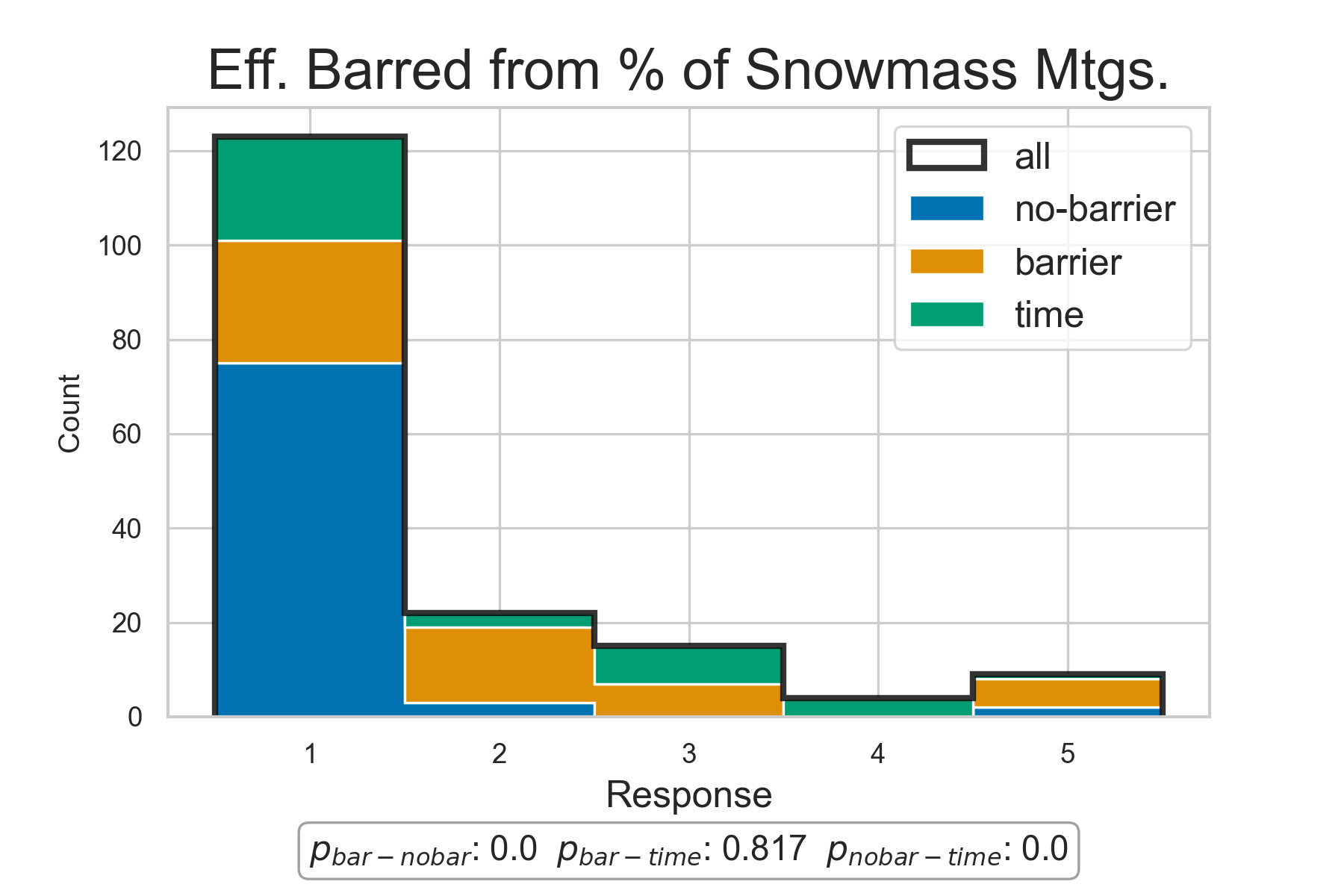}
    \caption{Question: What percentage of Snowmass meetings/activities have accessibility needs (if any) effectively barred you from participation?\\
    Response 1: $<$20\%\\ 
    Response 2: 20-40\%\\ 
    Response 3: 40-60\%\\ 
    Response 4: 60-80\%\\ 
    Response 5: $>$80\%}
    \label{fig:Q4_barrier_comparison3}
\end{figure}

\begin{figure}[ht]
    \centering
    \includegraphics[width=\textwidth]{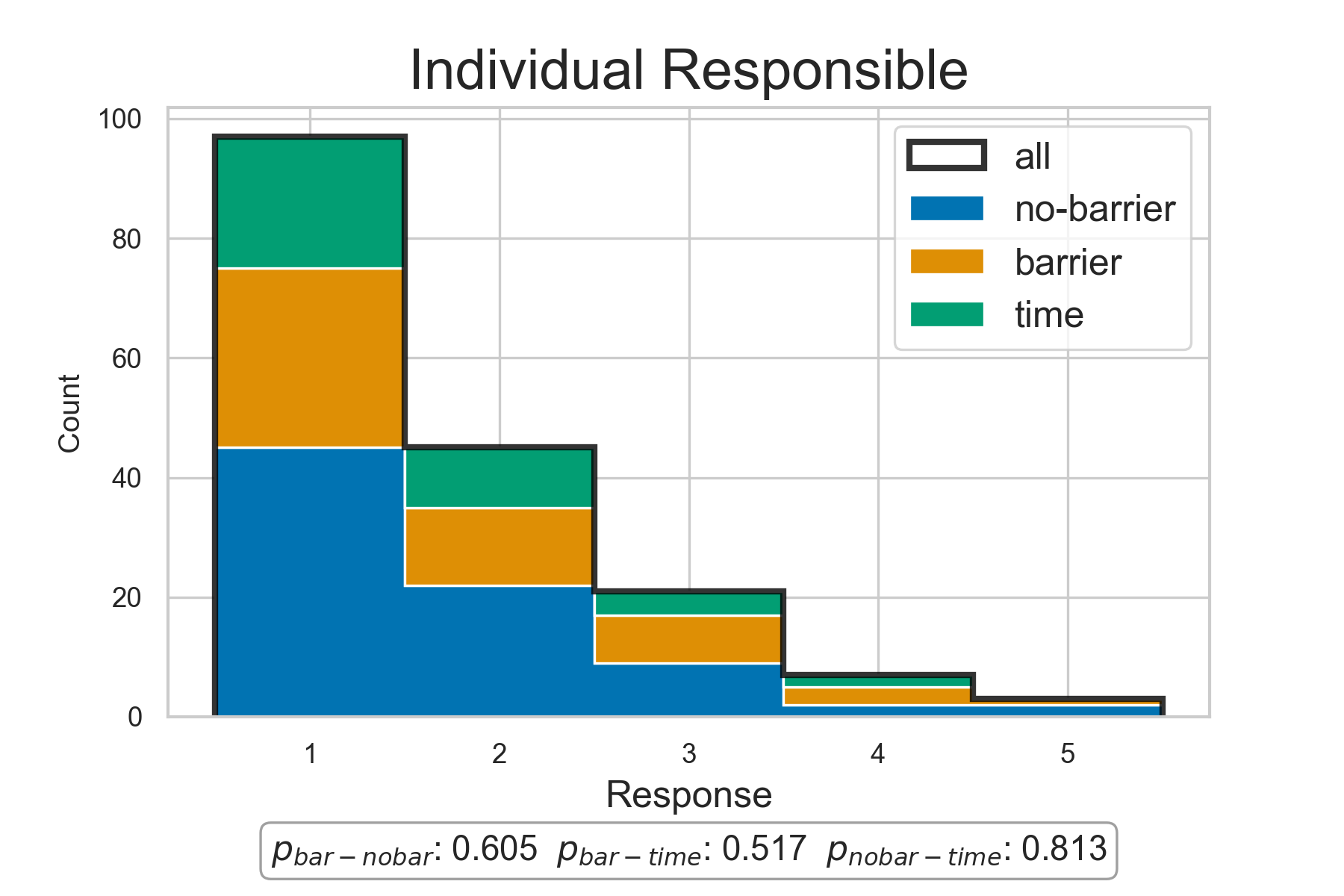}
    \caption{Statement: I feel that the financial responsibility for accessibility during \underline{US-based} conferences/community events should fall on the \textbf{Individual}.\\
    Response 1: Strongly disagree\\ 
    Response 2: Disagree\\ 
    Response 3: Neutral\\ 
    Response 4: Agree\\ 
    Response 5: Strongly agree}
    \label{fig:Q5_barrier_comparison3}
\end{figure}

\begin{figure}[ht]
    \centering
    \includegraphics[width=\textwidth]{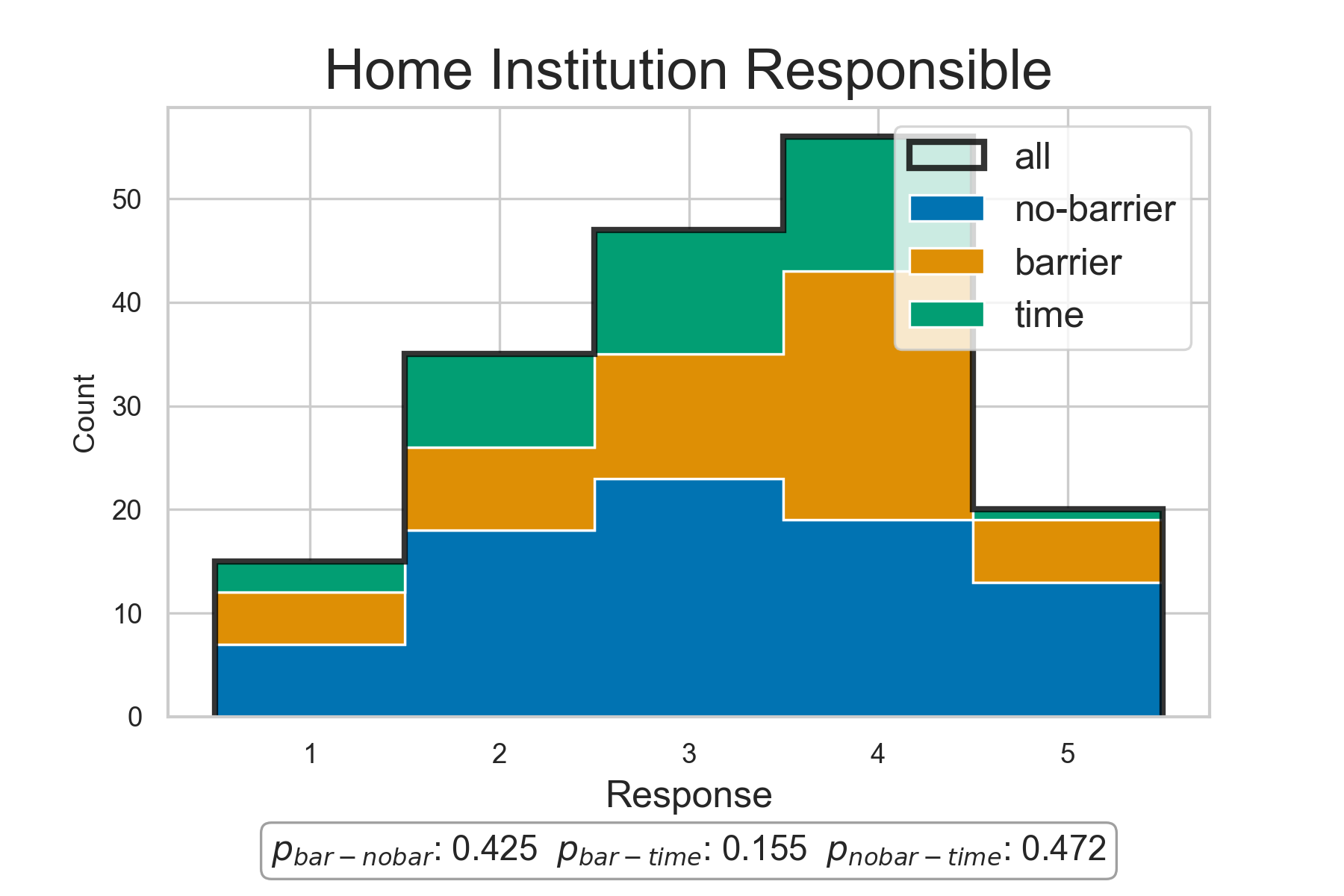}
    \caption{Statement: I feel that the financial responsibility for accessibility during \underline{US-based} conferences/community events should fall on the \textbf{Individual's Home Institution}.\\
    Response 1: Strongly disagree\\ 
    Response 2: Disagree\\ 
    Response 3: Neutral\\ 
    Response 4: Agree\\ 
    Response 5: Strongly agree}
    \label{fig:Q6_barrier_comparison3}
\end{figure}

\begin{figure}[ht]
    \centering
    \includegraphics[width=\textwidth]{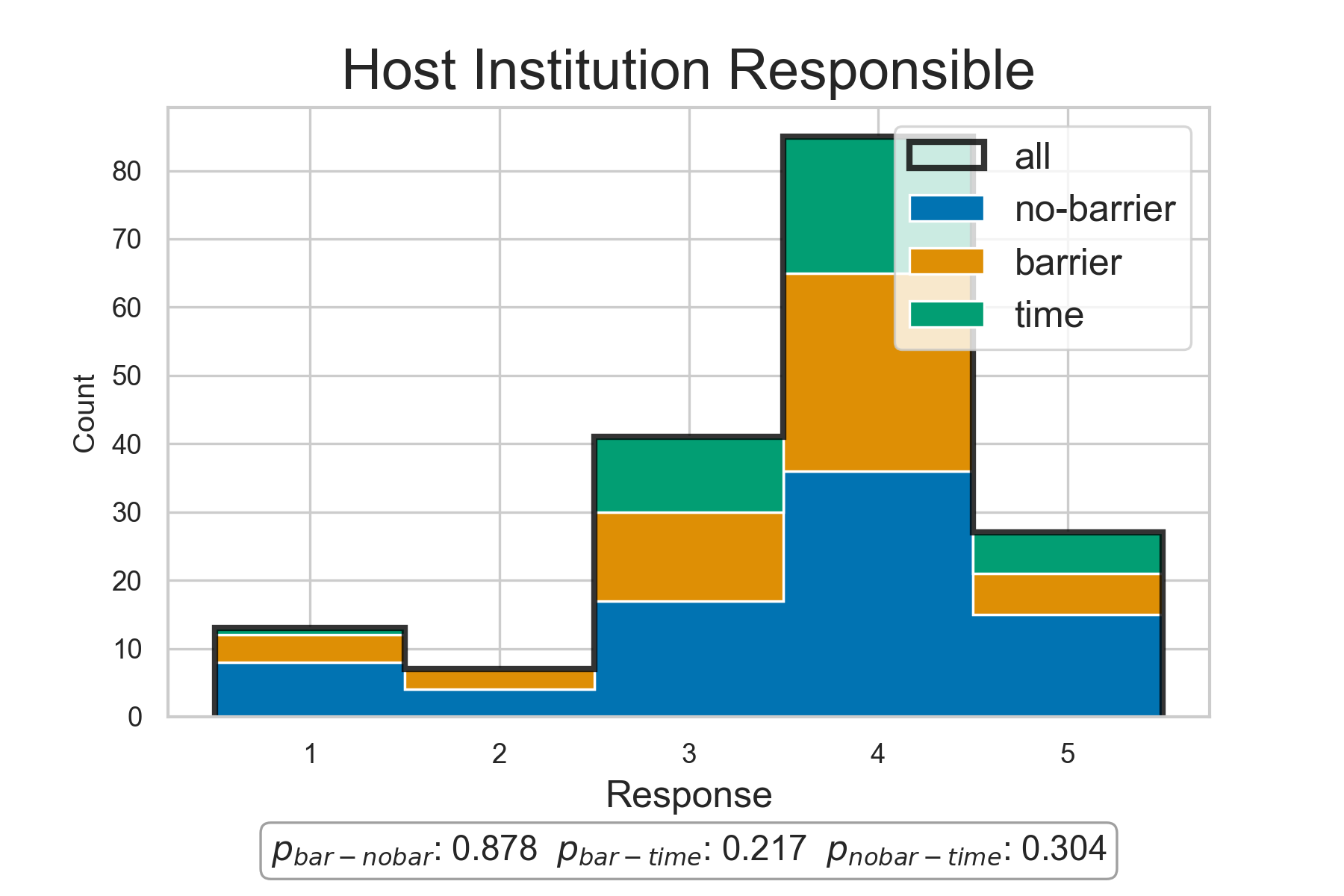}
    \caption{Statement: I feel that the financial responsibility for accessibility during \underline{US-based} conferences/community events should fall on the \textbf{Host Institution}.\\
    Response 1: Strongly disagree\\ 
    Response 2: Disagree\\ 
    Response 3: Neutral\\ 
    Response 4: Agree\\ 
    Response 5: Strongly agree}
    \label{fig:Q7_barrier_comparison3}
\end{figure}

\begin{figure}[ht]
    \centering
    \includegraphics[width=\textwidth]{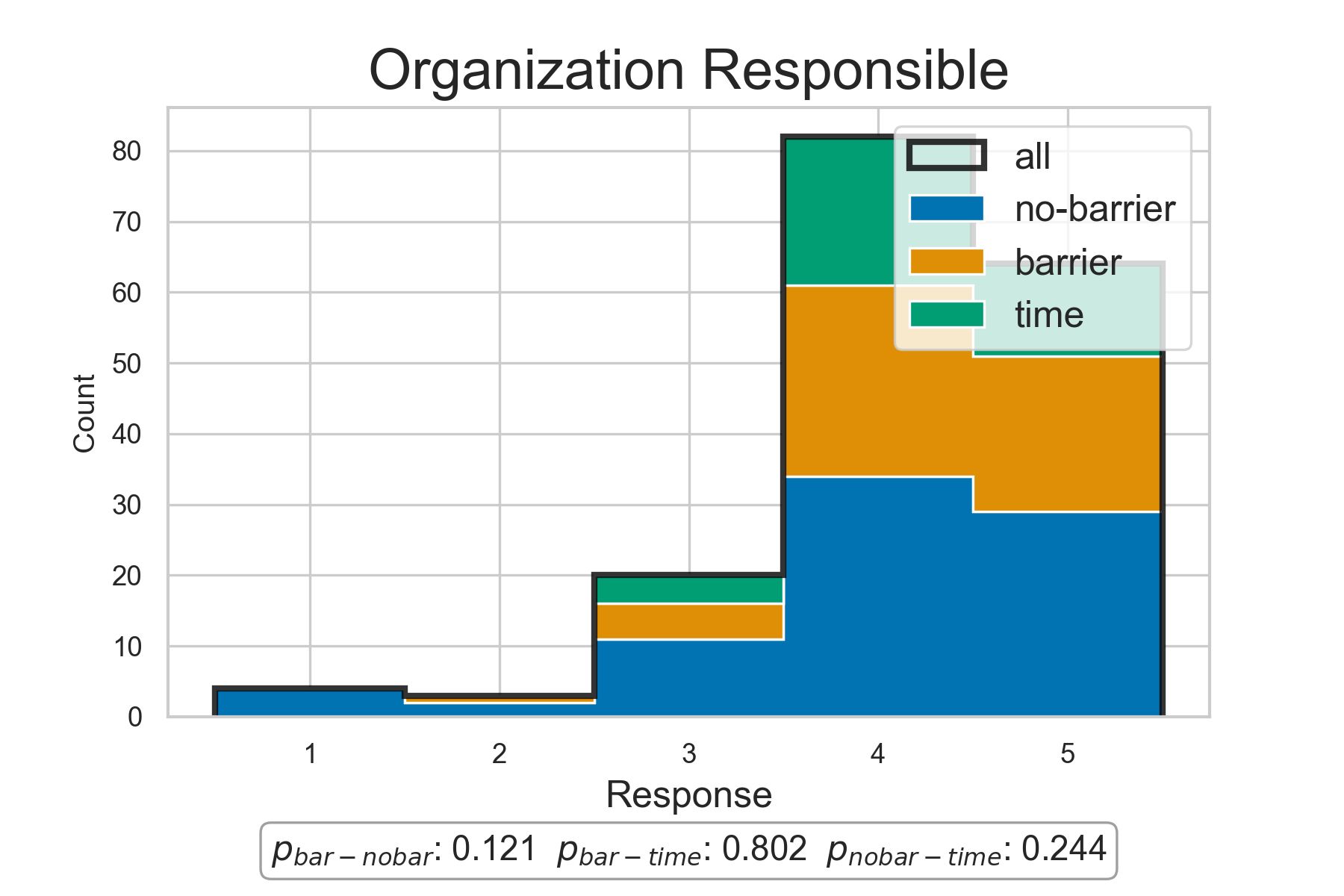}
    \caption{Statement: I feel that the financial responsibility for accessibility during \underline{US-based} conferences/community events should fall on the \textbf{Organizational Sponsors} (APS, DPF, etc).\\
    Response 1: Strongly disagree\\ 
    Response 2: Disagree\\ 
    Response 3: Neutral\\ 
    Response 4: Agree\\ 
    Response 5: Strongly agree}
    \label{fig:Q8_barrier_comparison3}
\end{figure}

\begin{figure}[ht]
    \centering
    \includegraphics[width=\textwidth]{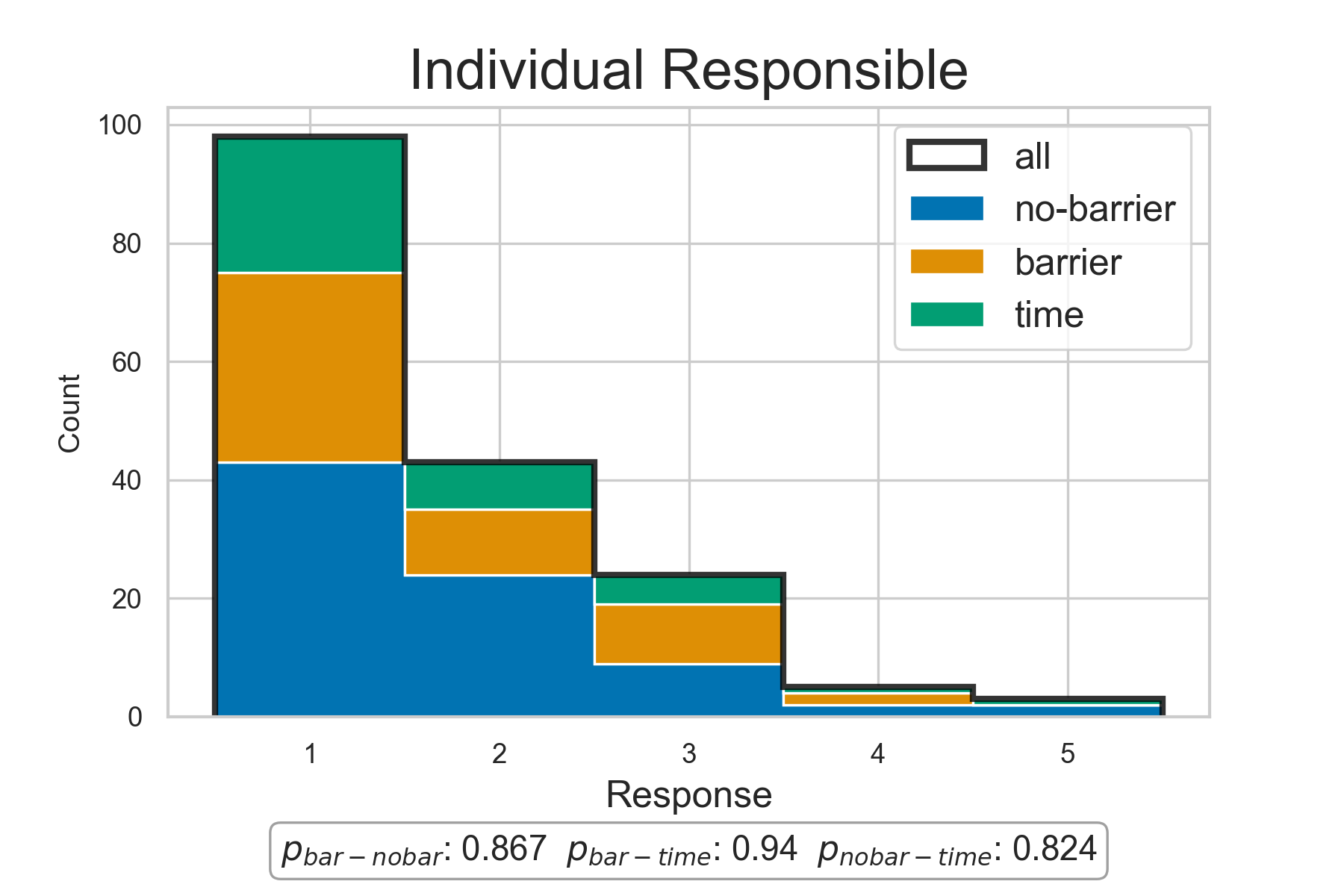}
    \caption{Statement: I feel that the financial responsibility for accessibility during \underline{international} conferences/community events should fall on the \textbf{Individual}.\\
    Response 1: Strongly disagree\\ 
    Response 2: Disagree\\ 
    Response 3: Neutral\\ 
    Response 4: Agree\\ 
    Response 5: Strongly agree}
    \label{fig:Q9_barrier_comparison3}
\end{figure}

\begin{figure}[ht]
    \centering
    \includegraphics[width=\textwidth]{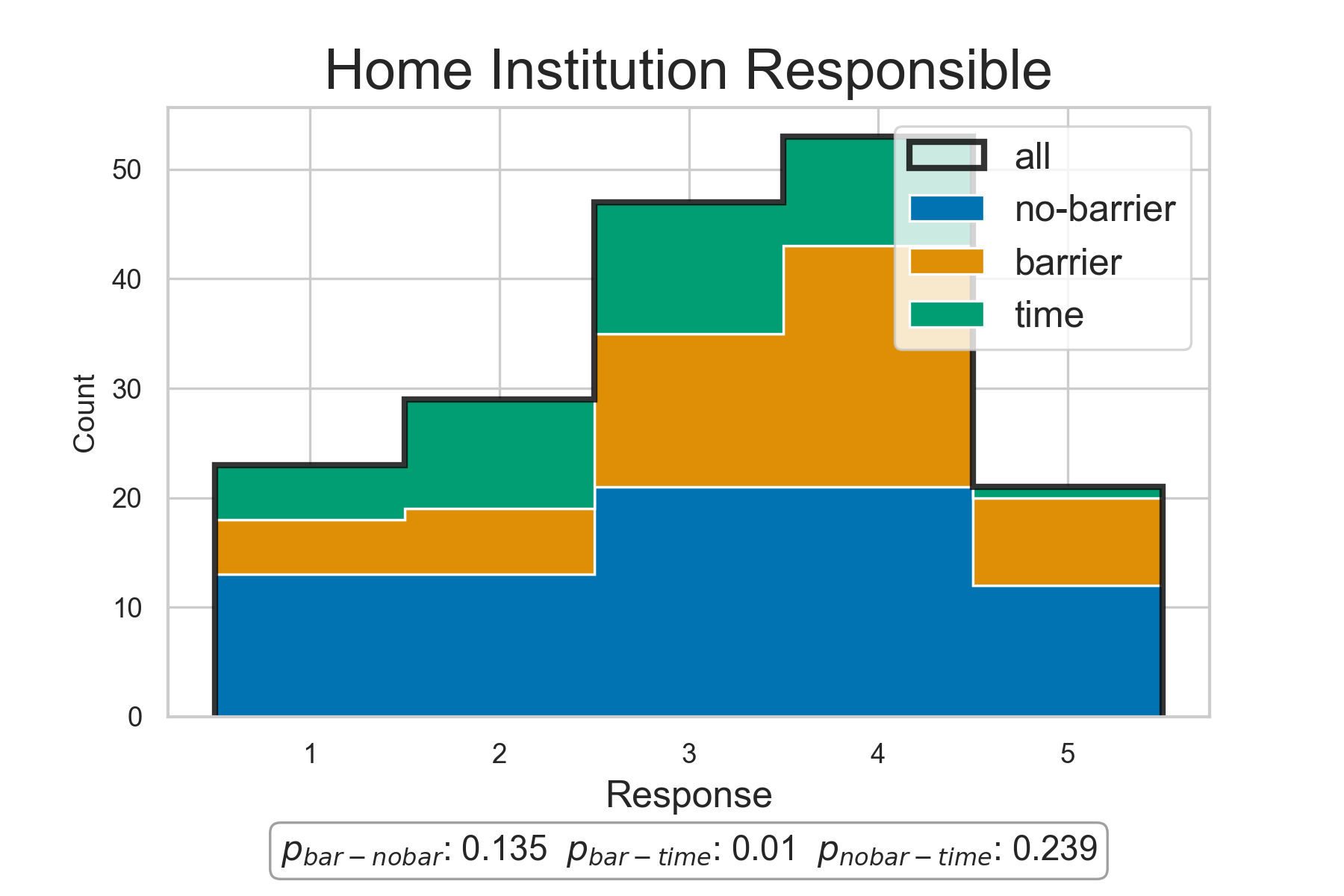}
    \caption{Statement: I feel that the financial responsibility for accessibility during \underline{international} conferences/community events should fall on the \textbf{Individual's Home Institution}.\\
    Response 1: Strongly disagree\\ 
    Response 2: Disagree\\ 
    Response 3: Neutral\\ 
    Response 4: Agree\\ 
    Response 5: Strongly agree}
    \label{fig:Q10_barrier_comparison3}
\end{figure}

\begin{figure}[ht]
    \centering
    \includegraphics[width=\textwidth]{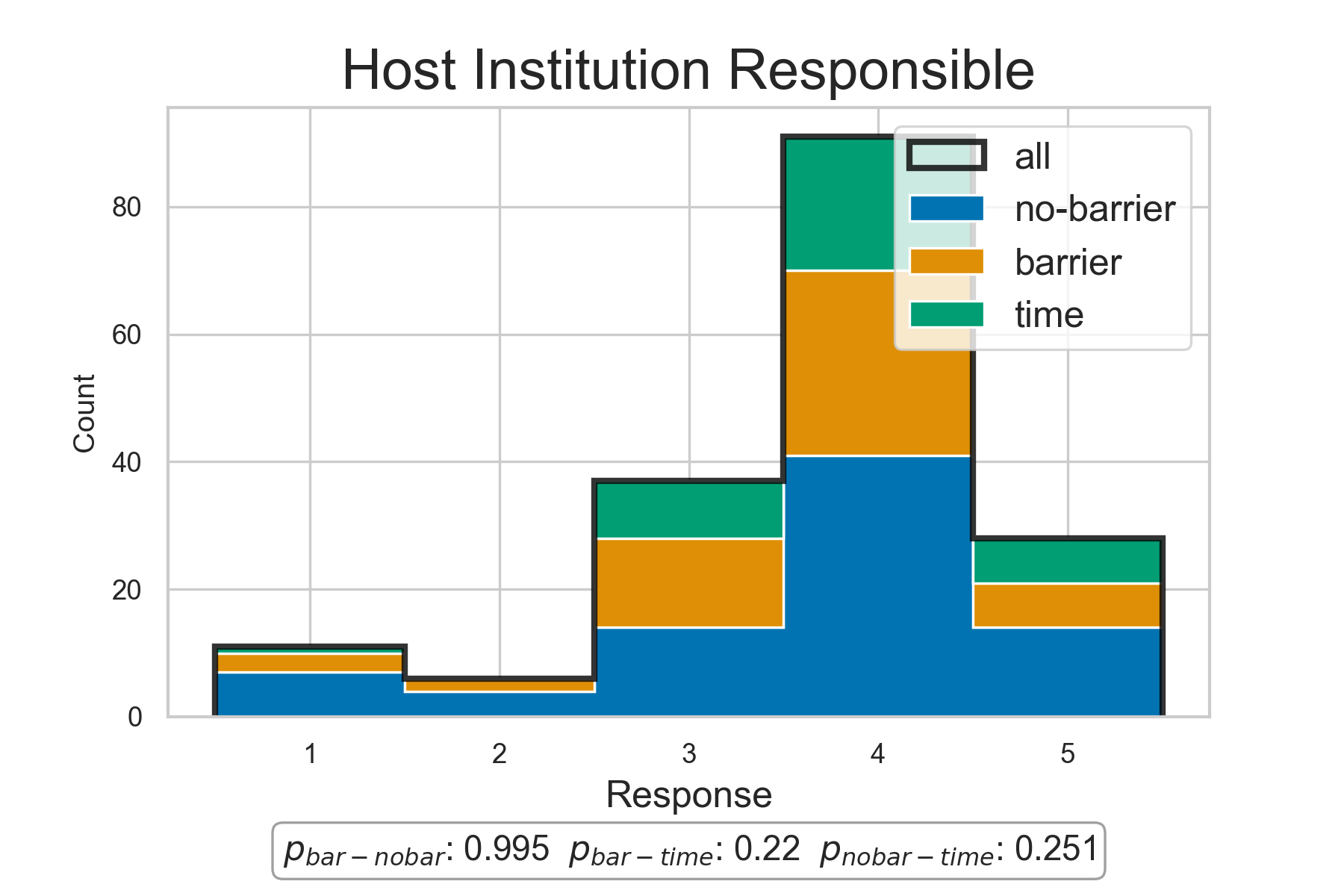}
    \caption{Statement: I feel that the financial responsibility for accessibility during \underline{international} conferences/community events should fall on the \textbf{Host Institution}.\\
    Response 1: Strongly disagree\\ 
    Response 2: Disagree\\ 
    Response 3: Neutral\\ 
    Response 4: Agree\\ 
    Response 5: Strongly agree}
    \label{fig:Q11_barrier_comparison3}
\end{figure}

\begin{figure}[ht]
    \centering
    \includegraphics[width=\textwidth]{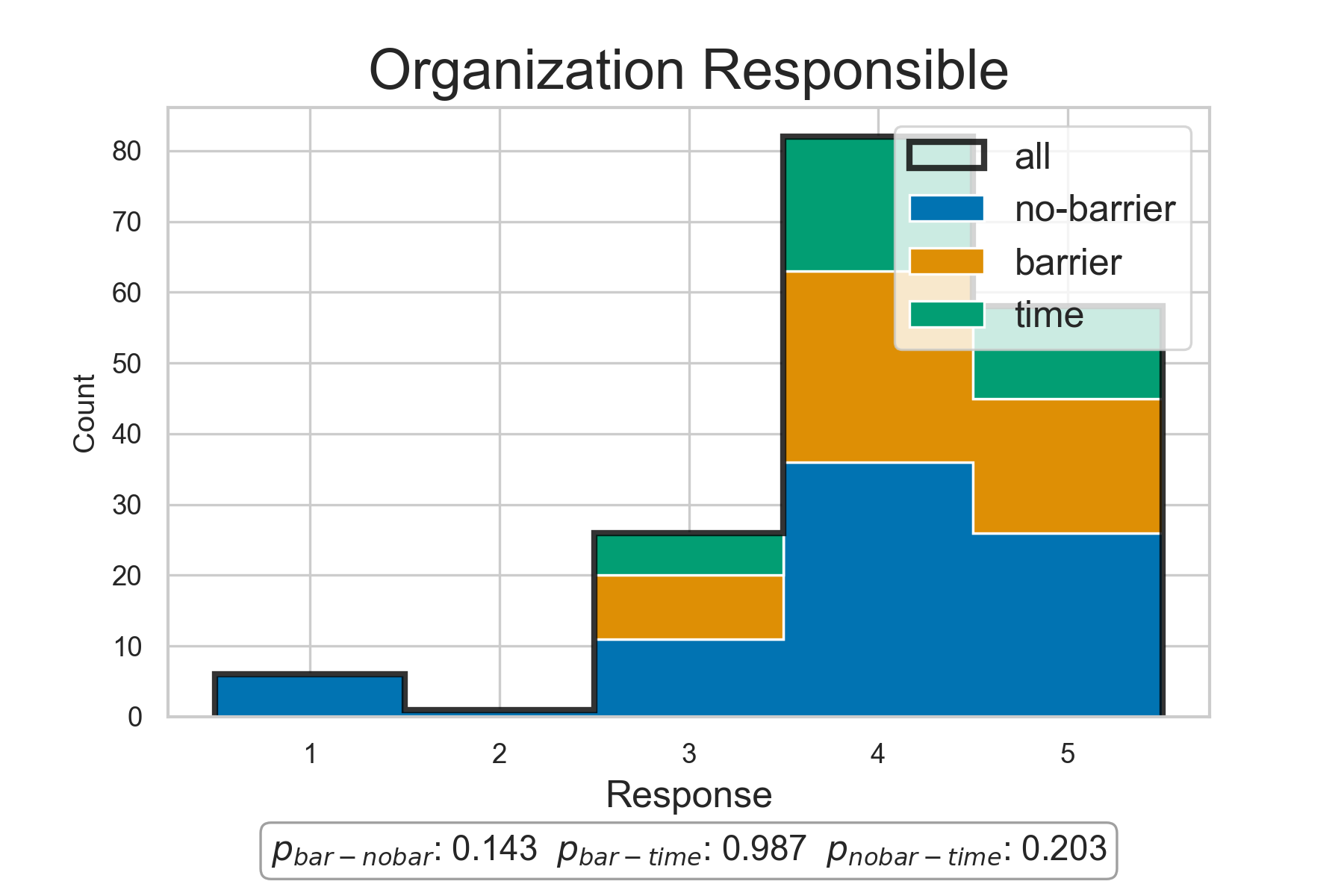}
    \caption{Statement: I feel that the financial responsibility for accessibility during \underline{international} conferences/community events should fall on the \textbf{Organizational Sponsors} (APS, DPF, etc).\\
    Response 1: Strongly disagree\\ 
    Response 2: Disagree\\ 
    Response 3: Neutral\\ 
    Response 4: Agree\\ 
    Response 5: Strongly agree}
    \label{fig:Q12_barrier_comparison3}
\end{figure}

\begin{figure}[ht]
    \centering
    \includegraphics[width=\textwidth]{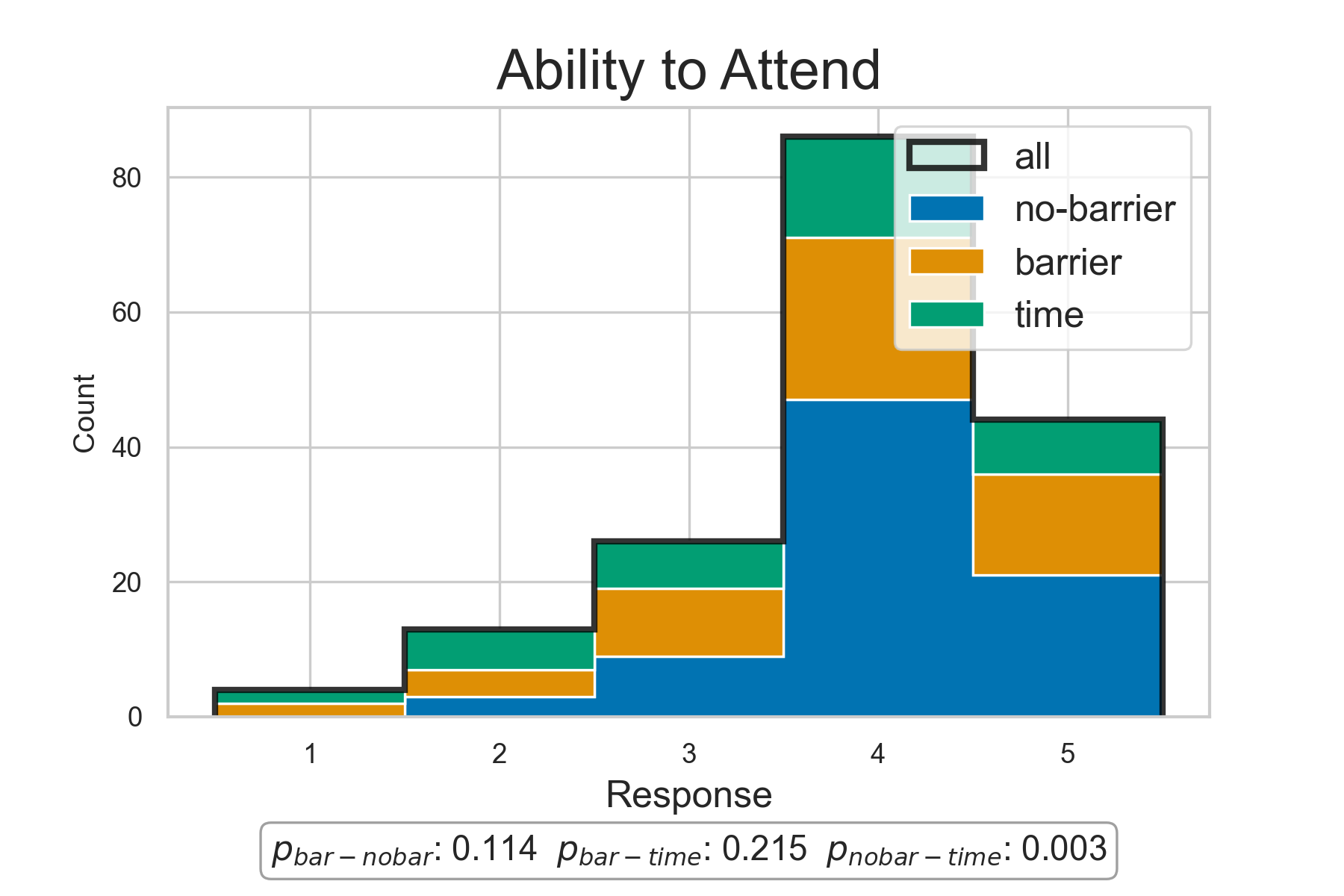}
    \caption{Question: What has been your personal experience with virtual conferences with respect to \textbf{Ability to attend}?\\
    Response 1: Awful\\ 
    Response 2: Poor\\ 
    Response 3: Neutral\\ 
    Response 4: Good\\ 
    Response 5: Amazing}
    \label{fig:Q14_barrier_comparison3}
\end{figure}

\begin{figure}[ht]
    \centering
    \includegraphics[width=\textwidth]{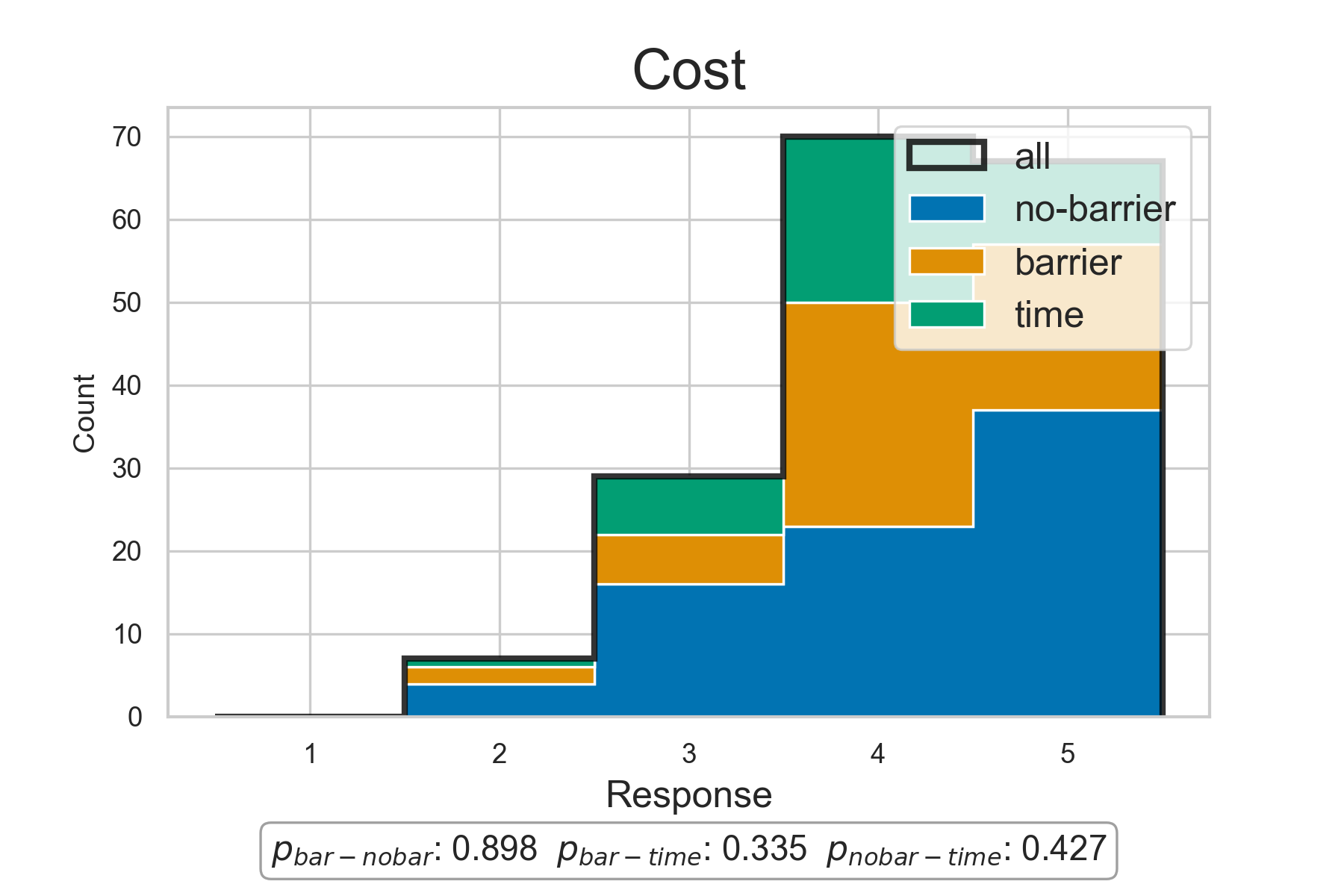}
    \caption{Question: What has been your personal experience with virtual conferences with respect to \textbf{Cost}?\\
    Response 1: Awful\\ 
    Response 2: Poor\\ 
    Response 3: Neutral\\ 
    Response 4: Good\\ 
    Response 5: Amazing}
    \label{fig:Q15_barrier_comparison3}
\end{figure}
    
\begin{figure}[ht]
    \centering
    \includegraphics[width=\textwidth]{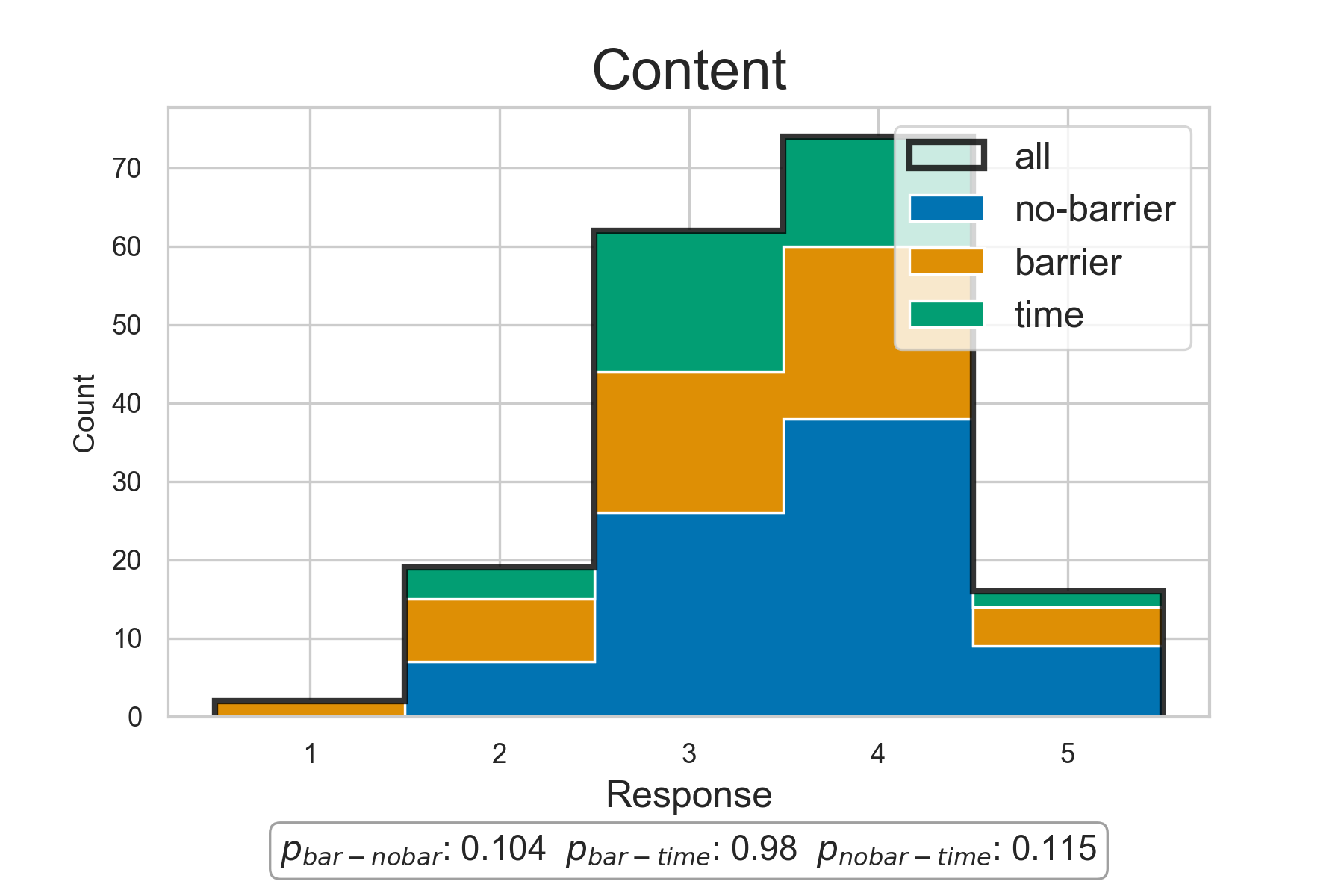}
    \caption{Question: What has been your personal experience with virtual conferences with respect to \textbf{Content}?\\
    Response 1: Awful\\ 
    Response 2: Poor\\ 
    Response 3: Neutral\\ 
    Response 4: Good\\ 
    Response 5: Amazing}
    \label{fig:Q16_barrier_comparison3}
\end{figure}

\begin{figure}[ht]
    \centering
    \includegraphics[width=\textwidth]{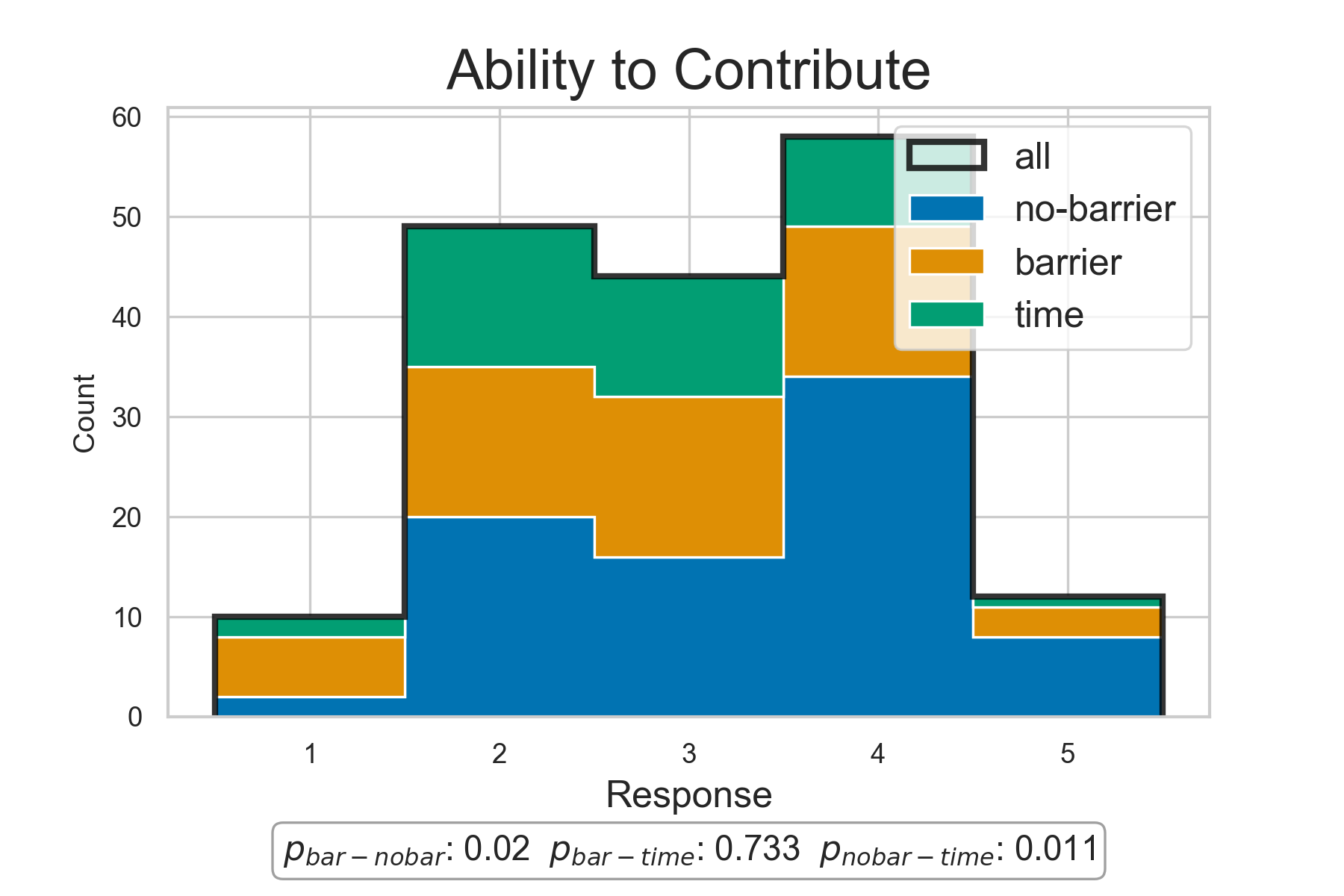}
    \caption{Question: What has been your personal experience with virtual conferences with respect to \textbf{Ability to contribute}?\\
    Response 1: Awful\\ 
    Response 2: Poor\\ 
    Response 3: Neutral\\ 
    Response 4: Good\\ 
    Response 5: Amazing}
    \label{fig:Q17_barrier_comparison3}
\end{figure}

\begin{figure}[ht]
    \centering
    \includegraphics[width=\textwidth]{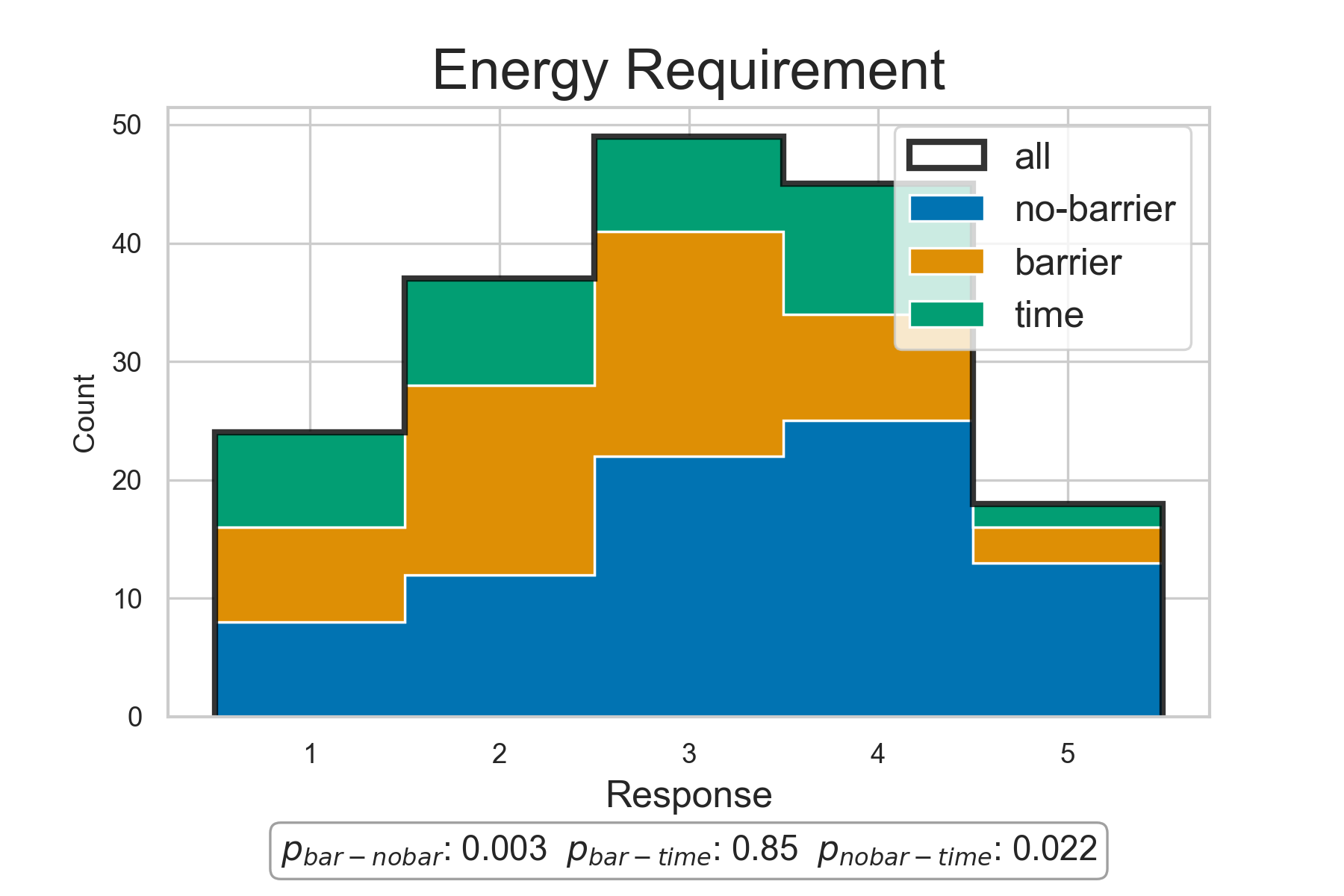}
    \caption{Question: What has been your personal experience with virtual conferences with respect to \textbf{Energy it takes to attend}?\\
    Response 1: Awful\\ 
    Response 2: Poor\\ 
    Response 3: Neutral\\ 
    Response 4: Good\\ 
    Response 5: Amazing}
    \label{fig:Q18_barrier_comparison3}
\end{figure}

\begin{figure}[ht]
    \centering
    \includegraphics[width=\textwidth]{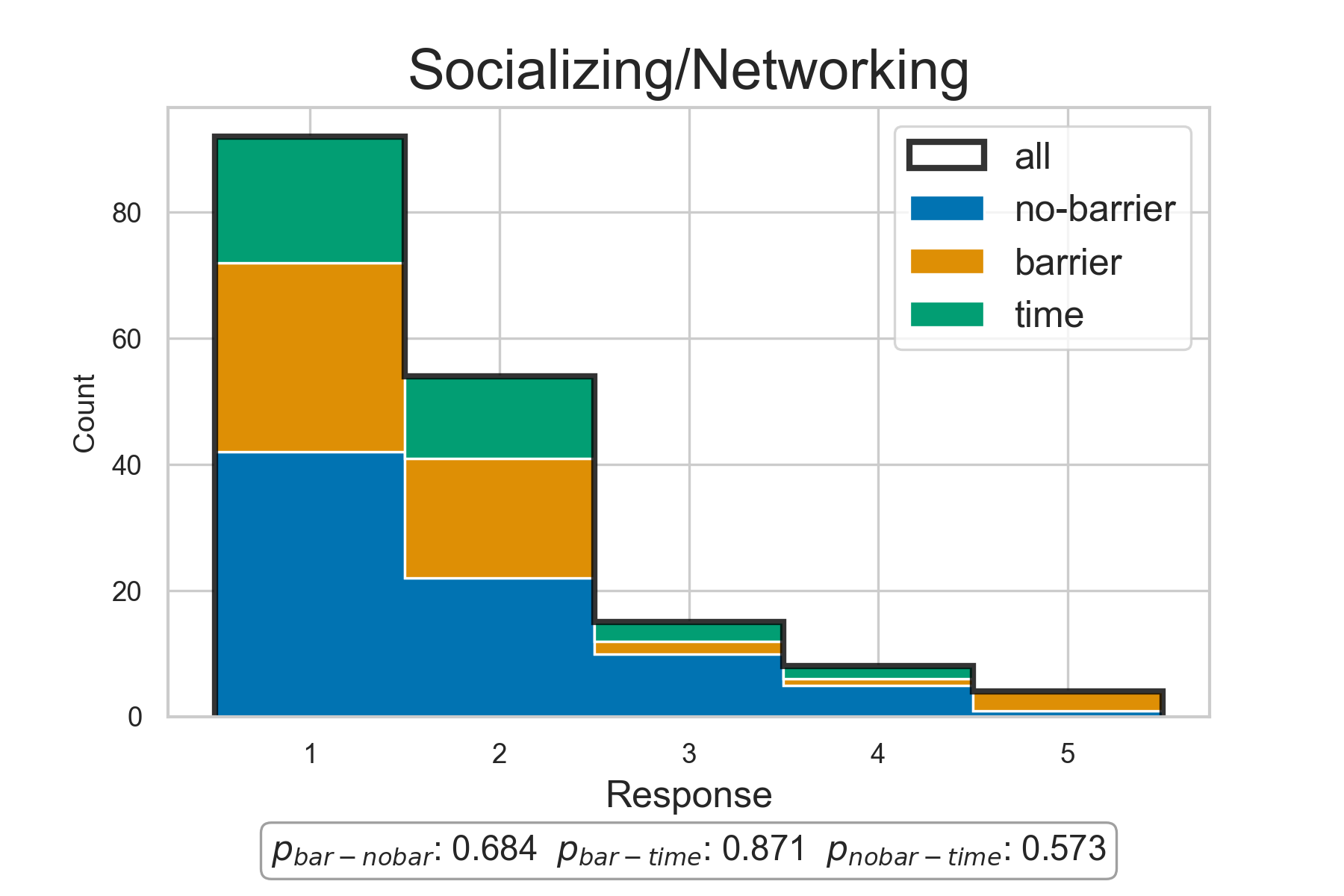}
    \caption{Question: What has been your personal experience with virtual conferences with respect to \textbf{Socializing/networking}?\\
    Response 1: Awful\\ 
    Response 2: Poor\\ 
    Response 3: Neutral\\ 
    Response 4: Good\\ 
    Response 5: Amazing}
    \label{fig:Q19_barrier_comparison3}
\end{figure}

\begin{figure}[ht]
    \centering
    \includegraphics[width=\textwidth]{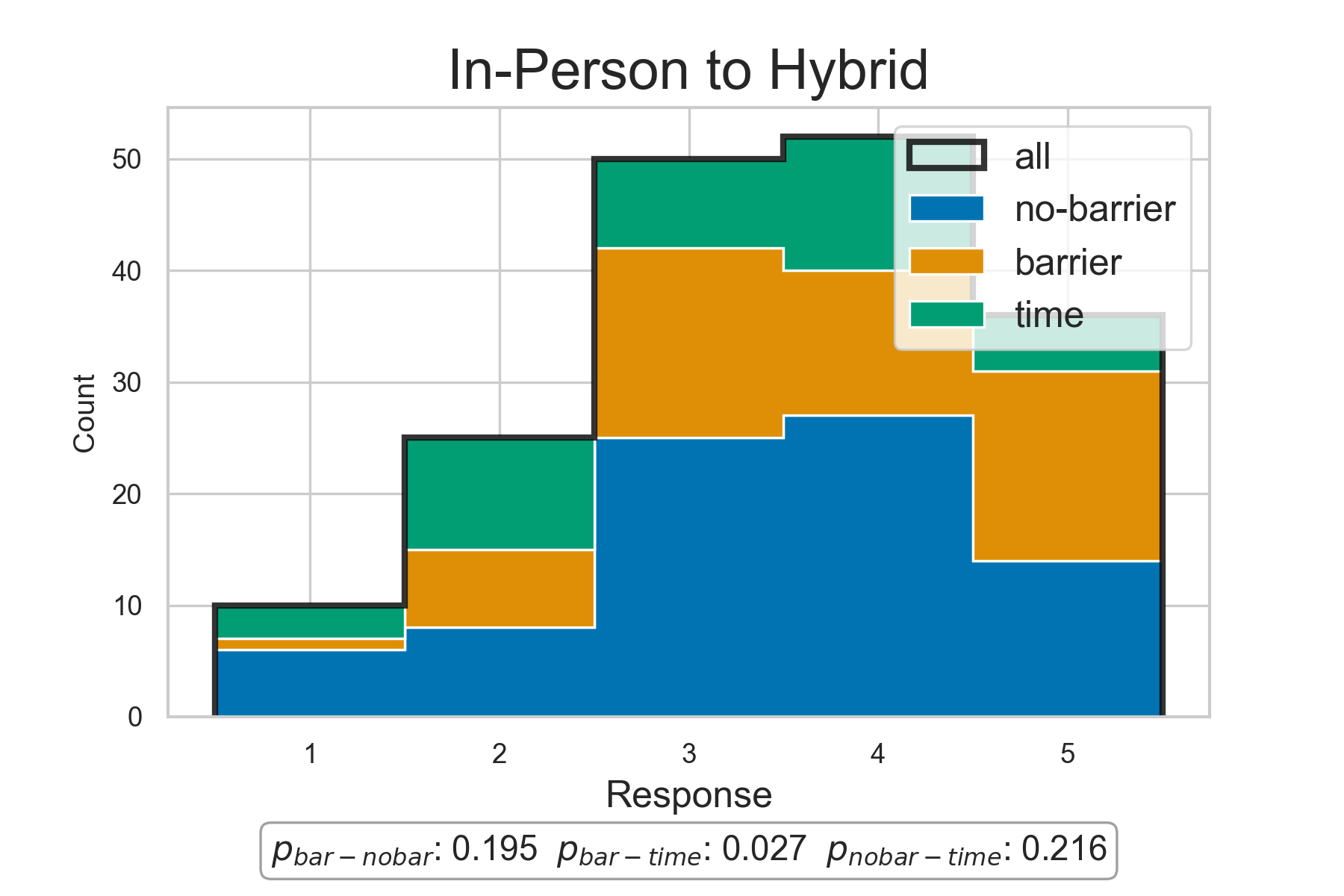}
    \caption{Question: If a conference is scheduled to be in-person, would you be more likely to attend if it were hybrid instead?\\
    Response 1: Very unlikely\\ 
    Response 2: Unlikely\\ 
    Response 3: Neutral\\ 
    Response 4: Likely\\ 
    Response 5: Very likely}
    \label{fig:Q21_barrier_comparison3}
\end{figure}

\begin{figure}[ht]
    \centering
    \includegraphics[width=\textwidth]{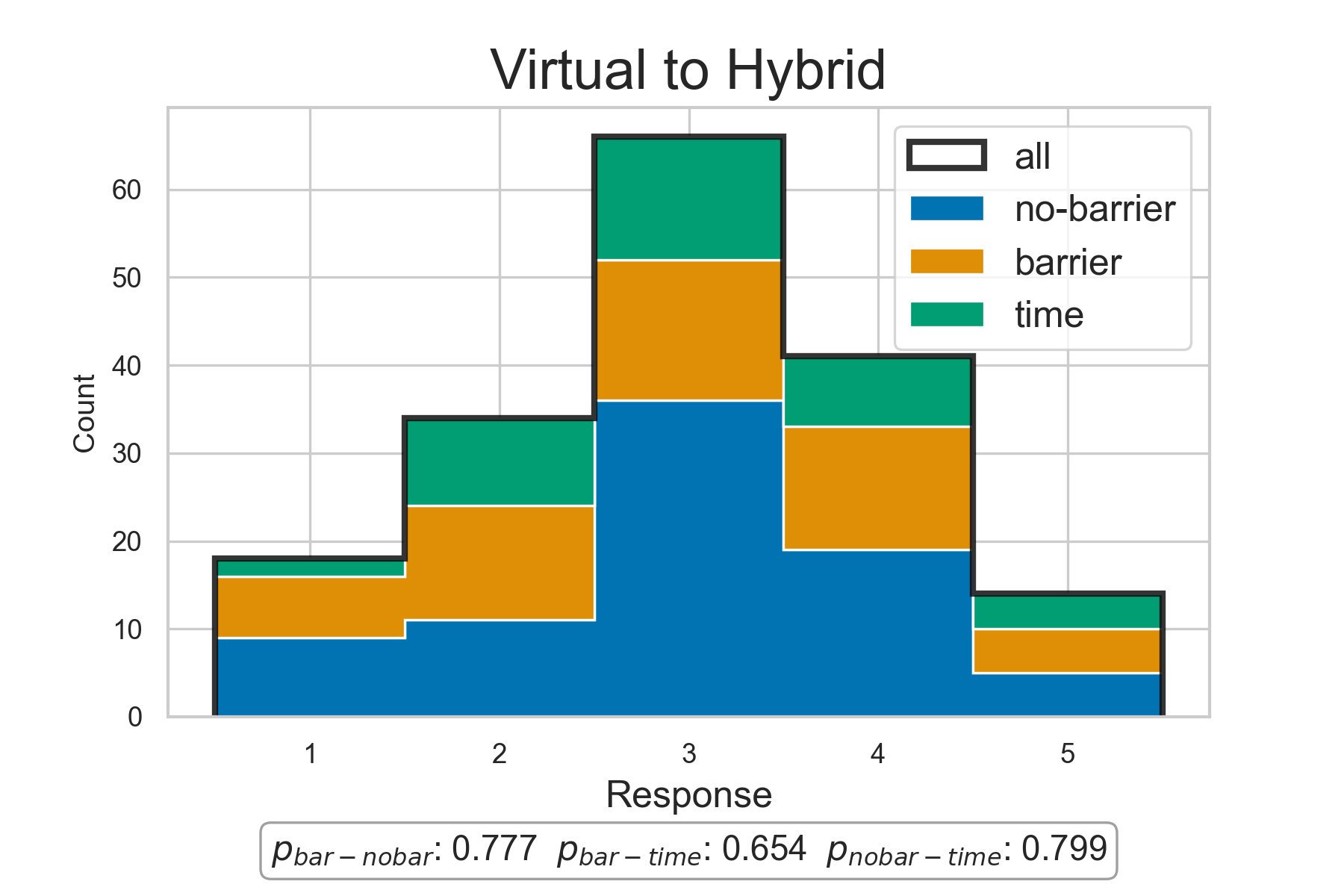}
    \caption{Question: If a conference is scheduled to be virtual, would you be more likely to attend if it were hybrid instead?\\
    Response 1: Very unlikely\\ 
    Response 2: Unlikely\\ 
    Response 3: Neutral\\ 
    Response 4: Likely\\ 
    Response 5: Very likely}
    \label{fig:Q22_barrier_comparison3}
\end{figure}

%\begin{figure}[ht]
%    \centering
%    \includegraphics[width=\textwidth]{fig/PLOT_NAME_GOES_HERE.png}
%    \caption{CAPTION_GOES_HERE}
%    \label{fig:IF_YOU_WANT_TO_NAME_THIS_FOR_REF_LATER_OTHERWISE_THIS_LINE_CAN_BE_DELETED}
%\end{figure}

%% file: tex/app_autocraption_comparison_susy.tex
\section{Comparison of YouTube Auto-caption vs Real-time Steno-captioning}

\label{app:autocraption_susy}

Within ATLAS, one of the collaboration members frequently has real-time \gls{steno}-captioning provided as part of workplace accommodations. A subset of an internal discussion was made public (unlisted) on YouTube in order to show the comparison between two different ways of providing real-time captioning for meetings~\cite{YouTubeSUSY2020}.

This appendix is formatted into two columns to allow the reader to understand the usefulness of auto-captioning (by \url{https://youtube.com/}; on the left, $\approx20\%$ \gls{WER}), with real-time \gls{steno}-captioning (by \gls{WCC}); on the right, $<1\%$ \gls{WER}).  Each block of text is displayed on the video at the corresponding timestamp. A few things are easy to notice:

\begin{itemize}
    \item autocaption cannot distinguish between speakers;
    \item steno-captioning understands where natural breaks in sentences are; and
    \item the lack of contextual awareness in autocaptioning, such as with capitalization, makes it much harder to understand what is discussed.
\end{itemize}

\begin{paracol}{2}
\footnotesize
\input{misc/app_autocraption_susy_youtube}
\switchcolumn
\raggedleft
\input{misc/app_autocraption_susy_wcc}
\end{paracol}

%% file: misc/app_autocraption_susy_youtube.tex
\SetLabelAlign{leftmargin}{\llap{#1\enspace}}
\SetLabelAlign{rightmargin}{\rlap{\hspace{\columnwidth}\enspace#1}}

\begin{center}
\Large YouTube auto-caption
\end{center}

\begin{description}[align=leftmargin,labelsep=0pt,leftmargin=*,font=\color{gray}]
\scriptsize
\item[00:00] to me uh as soon as
\item[00:04] we get this kind of discrepancies
\item[00:06] happening below a
\item[00:08] certain threshold that's when we can
\item[00:11] say make the automatic stitch from
\item[00:14] rude staff
\item[00:15] to pie a chef and i was wondering whether this
\item[00:19] is some
\item[00:20] maybe some numerical feature and whether
\item[00:23] you try to compute the play hf numbers
\item[00:27] with different backhands
\item[00:29] or whether this could be something like
\item[00:31] minnows versus
\item[00:33] uh um
\item[00:37] virtues i don't remember the name now
\item[00:38] yeah um
\item[00:40] i can comment so i think what's likely
\item[00:43] so
\item[00:43] like all the different back ends agree
\item[00:45] um so
\item[00:47] i think what's likely is that a um
\item[00:50] either
\item[00:50] so the numerical likelihood difference
\item[00:52] so the likelihood is only defined up to
\item[00:54] a constant
\item[00:55] factor right and so the way that um
\item[00:58] his fixture and rudes uh computes the
\item[01:00] likelihood
\item[01:01] and pie eff computes the likelihood is um
\item[01:04] kind of different by a constant factor
\item[01:07] because
\item[01:08] we literally invited to um compute
\item[01:11] let's say i've been by bin plus on and
\item[01:13] some it up over the events
\item[01:14] and we're root uh the sum um
\item[01:18] hacks with this histogram pdf
\item[01:21] and so i think so the numerical
\item[01:23] difference is a little bit different so
\item[01:24] the function that is being minimized
\item[01:26] is different by a constant factor and
\item[01:29] then also the minimization algorithm is
\item[01:31] different so i think
\item[01:32] it might be some confluence of these
\item[01:34] things so this i think has been the
\item[01:35] first time that we've seen
\item[01:37] a discrepancy on this level and so
\item[01:40] we have some ways to kind of offset this
\item[01:43] common factor and so that's
\item[01:44] how we're going to debug it lucas i
\item[01:47] think this is an excellent idea because
\item[01:49] it is very striking that
\item[01:51] these are two low mass points where we
\item[01:53] are potentially
\item[01:55] expecting rubber large needles which
\item[01:57] would translate
\item[01:58] in exactly this kind of offset problem
\item[02:01] and indeed
\item[02:02] when we are doing the fibs in history we
\item[02:05] are always using an offset subtraction
\item[02:17] okay are there any other questions
\item[02:23] [Music]
\item[02:35] um but um
\item[02:38] i've observed actually some deviations
\item[02:41] in the stop zero left um
\item[02:42] nothing they expected of observed but in
\item[02:45] the
\item[02:47] cls up with the
\end{description}

%% file: misc/app_autocraption_susy_wcc.tex
\SetLabelAlign{leftmargin}{\llap{#1\enspace}}
\SetLabelAlign{rightmargin}{\rlap{\hspace{\columnwidth}\enspace#1}}

\begin{center}
\Large Real-time \gls{steno}-caption
\end{center}

\begin{description}[align=rightmargin,labelsep=0pt,leftmargin=*,font=\color{gray}]
\footnotesize
\item[00:00] where to me as soon as we get this kind of
\item[00:05] discrepancies happening below a certain threshold,
that's when we can say make the automatic
\item[00:13] switch from RooStats to pyhf. And I was wondering
whether there is maybe some numerical feature
\item[00:21] and whether you tried to compute the pyhf
numbers with different backends or whether
\item[00:30] this could be something like Minuit versus
...
\item[00:33] I don't remember the name now.
\item[00:39] Lukas: Yeah. I can comment a bit. I think
what's likely ‑‑ all of the different
\item[00:44] backgrounds agree. So I think what's likely
is that, (a), either ‑‑ so the numerical
\newpage
\item[00:51] likelihood difference, the likelihood is only
defined up to a common factor, right? The
\item[00:57] way that HistFactory and ROOT computes the
likelihood and pyhf computes the likelihood
\item[01:03] is kind of different by a constant factor
because you literally in pyhf compute let's
\item[01:11] say a bin‑by‑bin Poisson and a sum over
the bins.
\item[01:15] And where ROOT, the sum stacks with the histogram
PDF. So I think ‑‑ so the numerical difference
\item[01:23] is a little bit different. The function that
is being minimized is different by a constant
\item[01:28] factor, and the minimization routine is also different.
So I think it may be some combination of these
\item[01:34] things. So I think it's the first time we
have seen a discrepancy on this level.
\item[01:39] We have a way to sort of offset this constant
factor and that's how we're going to debug it.
\item[01:46] Jeanette: Lukas, I think these are next
\item[01:48] in the ideas. Because it's striking, these
are two mass points where we're expecting
\item[01:56] rather large news which could translate this
kind of offset problem. And indeed when we
\item[02:02] are doing the fits in HistFitter, we are also
using an offset construction.
\item[02:07] Giordon: Okay. Are there any other questions?\\
Krzysztof: Concerning this, I mean, validation
\item[02:25] because as far as in the strips here or the
CLs where ‑‑ I guess on this CLs where
\item[02:30] they were checked, but I observed some deviations
in the top 0 lepton, nothing that was expected
\item[02:34] to be observed, but in the CLs up or down
in expected. So I also want to check other
\item[02:45] CLs values or what's it take on this?
\end{description}

%% file: tex/app_autocraption_comparison_ef.tex
\section{Comparison of Otter.AI vs Note-taking}

\label{app:autocraption_ef}

Within one of the Snowmass meetings, community members of the \gls{EF} tried out Otter.ai. After some technical glitches, they managed to enable it, and part of the outcome is shown below on the left. Because there was someone who was attending the meeting and took notes, you can compare to roughly what was said.

This appendix is formatted into two columns to allow the reader to understand the usefulness of auto-captioning (by \url{https://otter.ai/}; on the left), by providing important context through live notes taken by physicists (on the right). This took time and effort on their behalf during the meeting, as well as additional time after the meeting to clean up their typos and re-summarize the discussions into minutes for the meeting. Names are abbreviated to initials to provide a level of anonymity, although these notes are publicly available. Horizontal lines are included to separate the different questions and discussion topics that occurred to help with legibility and try to understand what the automated transcription provided.

In particular, a few key \textsl{features} made the transcription pretty hard to follow, including:

\begin{itemize}
\item speaker accents and audio quality;
\item high use of acronyms, jargon, and other contextual clues;
\item lack of ability to use vocabulary, train the AI, or edit in real-time; and
\item no oversight on the quality.
\end{itemize}

\begin{paracol}{2}
\footnotesize
\input{misc/app_autocraption_ef_otter.ai}
\switchcolumn
\raggedleft
\input{misc/app_autocraption_ef_human_notetaking}
\end{paracol}

%% file: misc/app_autocraption_ef_otter.ai.tex
\SetLabelAlign{leftmargin}{\llap{#1\enspace}}
\SetLabelAlign{rightmargin}{\rlap{\hspace{\columnwidth}\enspace#1}}

\begin{center}
\Large Otter.AI Transcription
\end{center}

\begin{description}[align=leftmargin,labelsep=0pt,leftmargin=*,font=\color{gray}]
\scriptsize
\item[23:28] Yes, thank you. I want to comment on something that both yulian Tao, and Catarina mention, which is when we talk about multiplets. So you guys have a good amount of emphasis both on the casino and we know that coming from sushi. But, of course, other multiplets that may be less motivated from, from a modeling point of view, but they also useful to look for them. I'm talking about the minimal Dark Matter idea of what is multiplayer that that fulfills of the conditions. You have the web to exceed another too big, which is that we know about the other possibilities but it would be good also to have benchmarks on understudies from on on those scenarios. Thank you.
\item[24:09] Yeah, exactly. So, so I do have many more clutter on my slides as well so I think that he know we know are just studying for doubles and triples. So we could certainly have more multiplies yeah that's a very good good suggestion.
\item[24:35] So, yeah.
\item[24:40] Do you know i mean Antonio Do you know that. What's the main difference between the mean, I think the cross section is different and in terms of signal what was the main difference. Well, you're right the the cross section are three different.
\item[24:56] The main thing now is, once you go beyond the triplet thing you end up having objects with the charge two or four or so on. So, that is a different scenario for example, I the charge to live long enough for you to tell.
\item[25:17] I forgot tiene TV saying something in the, in the church and there are also studies with meteors Of course, attributed to debates in Britain to WGS a singlet and these are well temper scenarios.
\item[25:34]
Yes, they, I don't know on top of my head and tell but certainly once you go beyond the targets and the chief that you, you start having double charge objects that are very interesting and of course as you mentioned, and also, I think Catarina mentioned, we always have on the back of our mind that the W and the T plus r are fermions because that's what happened in sushi. But of course these can be also scalar for example the scalar web. I think you get the right drug, correct me if I'm wrong and dogs around 500 gV s and E.
\item[26:08]
Yeah, I think, yes. Obviously, change the benchmark starts to change.
\item[26:18]
And, yeah, so, in terms of mixtures like, you know, more like a single entry tablet and Tim i think is making this point in the chat.
\item[26:29]
Yeah, so so these are becoming lines, instead of points. So yeah, so I think all of these. All of these are very interesting and I think we should certainly take all of them. Yes. Okay.
\item[26:49]
Cow your your your your hand raised and lowered. So yeah, I just want to say that typically if you go with the higher dimension higher cage representations. You lead to higher dimension operators, with several Higgs fields to form casing around operators. So, they are studies being done. So typically the signature to be be rather different however there are many more parameters, because this is not not even lower dimension operators.
\item[27:20]
Yes, in terms of their coupling to fix. Yes, and.
\item[27:26]
But in terms of their simple coupling to W and z that these are determined by by by just engaging gauge simple gauge interactions. So I think the production is still pretty standard. But in terms of their their associated signatures will say, Thank You are absolutely right.
\item[27:44] Yeah. I'm sorry Tim suck Yo yo yo yo Raise your hand. Yeah, raise my hand but then you Intel both said everything that I wanted to say so you could. Oh, sorry.
\item[28:00] Yeah, yes Sabina Sabina is pointing out something also very interesting is, whether you have Dirac or my urine out yet. Yeah. So, yes, this is another way of, you know, extending the minimal Suzy maker derocker.
\item[28:25] Right, so at least in the, in the
\item[28:30] US, at least I can imagine I, you know, in a simple mediator models that the weather dark matter is Iraq or Marana can have qualitative difference in terms of channel media turn models and.
\item[28:46] Yeah, so, but but you know certainly know tell us more about you know what you have in mind, you know, what kind of, I think the important thing for for Snowmass is, you know, it's usually not a complete exploration of model space but to come up with interesting benchmarks and, you know, God guided goalpost exploration.
\item[29:28] Any anybody ever so so far I think the comment has been mostly more like a
\item[29:37] simple web site, and anybody has any comments on you know more acceptor models and relating to their relation with dark matter.
\vspace{0.2cm}
\hrule
\newpage
\item[29:56] Well if you lead me oh sorry as it's wrong who has a question. Yeah.
\item[30:02] I would probably yeah so on Dantonio.
\item[30:06] Okay so, because this is my first time I'm not trying to pro exactly what is proper to be added, and advice not really proper to the editor but nonetheless, given the fact, this is my first time then we just say what I want I can say, so often right document or models, forgetting about the students, the first, second is organized according to, first of all, whether documentaries are charged understand the model, these are really these when we know you know that directions. And if they're not charged to the end we talked about the portals at the dimension last or equal to four, or we talked about hidden values if he's higher dimensional. And the question is, so So, most of the topics that are covered today. I believe are within these categories, right. So we talked about util ramp and then we talked about his portal, we talked about photon that photon, and then presumably we can talk about the trainer portal as well.
\item[31:06] Yes. And I wonder sample. This is the full.
\item[31:13] You know list or just general possible ways to tunnel into or probe into the Dark Sector. And again, we wonder about you. Do you have any. Yeah, for example.
\item[31:32] Yeah, if, if I can like share or share my ideas, which is not like put out yet, then I'm happy to say but I'm not sure if that is the kind of mode that we are we are doing this discussion, though, is that okay to for example throw us some random thoughts or it's more about the given work that thought and maybe trying to collect those. Well I think I think you can you can say it, but perhaps more importantly you can you can give us input on writing it down in the notes and also tell us you're interested in working on it. Yeah, okay. So, I mean it's underway right now for example that but.
\item[32:14] So, so far, you see the either document are charged, or just through the portal. This is all based on.
\item[32:22] If you reach local symmetries meaning by local symmetries I mean, on a local aspect of the field theories. And so you just write down, Jason var in local operators and that asks for its consequences at like different frontiers of experiments right now, what are the thoughts that they have, which are under exploration is that there is a rather slightly more global reasons for the tech sector to cough up to the standard model, and not a list some way I thought this is in a different category, compared to say, or in contrast to say either portal has a Hidden Valley. And in the title committer. And so the rich, then the concretely what I'm thinking of is basically dr capaz de Sena model do to say someone normally reasons, there's a shared anomaly. And we note some examples that if there is a sharing of the normal sector in the low, low energy effective theory, there can be certain interaction induce famous famously zoom in or we could be center or desktop, for example pi and Caicos the photon, and whatnot. So, I do have a model right now on the work but I can say anything concretely, whether you can program that, but anyway so yeah thanks yeah please do let us know. You know, if you are interested in writing input to the syllabus. Yeah, yeah. Okay Antonia.
\vspace{0.2cm}
\hrule
\newpage
\item[33:51] Yes, another more thing that we could think about is, is beyond the minimal scenario of of dark matter of minimal dark matter is when you have coordination with other particles, again, yes I'm being guided by the way we're having sushi. Yeah, it could be a situation with a simplified model that you have a singlet and color particles. It doesn't have to be ugly no but you know who you know where I'm going. Yeah, exactly. So, so if you look at my slide you know there are a lot of There are also already several Susy motivated the corner nation scenarios included there was, you know, both colored particle and also Urbino, you know, correlation so but yes, I think, in general, these are, give us additional lines, if you want, yeah. In addition to two points, we should be thinking about. Yes.
\item[34:55] So I'm not reading as fast as the chat window anybody wants to speak up about what you typed in the chat with no
\item[35:06] okay do you die, well maybe I will take a question by you die. Yeah. Yes, thing just Yes, absolutely completely updated like the dark photon plot. So yeah, so he, in the future, there are summary plot I can volunteer to help keep it updated. I'm most are not familiar with everything but I can get a sense of what's the most updated constraint for a lot of these mediator, like skater mediator, a mediator or having a lot of talk and help getting the plots updated for future summary plots I think this would be really good. And this is something that has, we'll have to discuss with the rare and precision.
\item[35:52] Dark Sector subgroup that got in touch with us but can't be here, at least one of them can be here because Australia. But then I think the foreigner is here. Yeah. And, yeah, I think.
\item[36:04] Yeah, we'll decide for the for the acceptors sushi sir.
\item[36:09] We'd be as far as the Send on in coordination with our group. So our, our group focuses on rare processes at a low energy and this includes that center central energy. So, there will be certainly some, some work done in, in collaboration with this group. So, in principle, you will know those a you know a tip or maybe several of those future bi weekly meetings where we can, you know, focus more on these topics. I think this is a very important topic. Yeah, absolutely. Yeah, yeah, yeah. So I just want to say. I think if I didn't, if I pronounced correctly Yeah, Philip Beaton like there's already several people that they have a compliation of this kind of constraints that they've published openly, so we can also use, and also directly get feedback from them to see, to get all of these updated. Actually I think he's here. Yeah, yeah I'm connected. So yes, we can we can do that.
\item[37:16] Do I pronounce it correctly. I've eaten. It's Elton Elton sorry. Yeah.
\item[37:24] Not Not a word.
\item[37:29] Fix.
\item[37:42] There was a more or less parallel conversation on the chat about the non tree level mediation between the standard model and dark matter then I want to bring it to voice.
\item[37:56] Maybe I can say something this is Tim, and unfortunately I've got to leave actually in a couple of minutes for another meeting, but what I had in mind was models where there's actually no tree level mediation between the dark matter and the standard model and so everything is intrinsically loop level, which is somewhat different than what Leon Tao picked up on which is of course that they're also often very important level corrections, even the processes that do have tree level pieces. So, these. If the mediation is sort of intrinsically not leading order there's a very interesting momentum dependence that's very different than the simple scaling you get from tree level models. And so that's something that's worth, you know it's sort of a corner maybe if you want a theory space but it's something that's worth looking at because it's rather different.
\item[38:41] Yeah, so, DC studies mostly focused on direct detections. No, I'm talking.
\item[38:48] For example, oh, okay, I put a link in the, in the fanciest no no I see it yeah sorry yes. Sounds good.
\item[38:58] Okay, yeah.
\vspace{0.2cm}
\hrule
\newpage
\item[39:08] I have a question to the convenience convenience Hi this is Marcus, um, do you have a sense of what the ratio of theorists, and experimentalists is in this working group.
\item[39:23] I don't know. I mean, we've been seeing emails come in so that is possible that, that is our.
\item[39:31] At least the people that I know for probably mostly experimentalists and I wouldn't say there's a big, big difference there's like no theories and all experimentalists I would say that there is a good split for now.
\item[39:45] Yeah, there are, I think it's probably a pretty good split.
\vspace{0.2cm}
\hrule
\newpage
\item[39:52] Okay, thank you.
\item[40:02] So, by the way, I think Jonathan also asked a question on the, on the life notes about like certain secrets mostly mainly about lonelier particles. Right.
\item[40:14] Yeah, and I did this reading, I think.
\item[40:18] No, this is about to about to end, I think, hmm, sorry I got the hour wrong. Yeah, sorry yeah the time shows on the Indiegogo page is the is the central time. So, I think.
\item[40:33] Yeah, the lovely particles, I think the topic itself, officially is managed by by EF oh nine but of course we have a lot of overlap with, with many other many, a lot of overlaps. So, so we'll we'll try to talk to each other as much as possible, and then we will be focusing on in the dark matter is that is the document or connection with the longer particles. So, these are strictly speaking to you know the not the London particle model.
\item[41:07] Full model space per se but the document or connection.
\item[41:11] Okay good, that's good. So I had a little correspondence with rare processes six.
\item[41:18] Mike Williams, and Stefania Gauri. And I guess they're also involved with this area. Are you coordinating with them as well. Yes yes Stefan yeah is here actually. We had someone email exchange was done we're probably going to even have some joint discussion together so we're going to have some two week bi weekly meeting about, you know, various go into more detail topics, perhaps or, you know, one of those meeting we'll be joined today with. So, so yeah you're certainly very much encouraged to attend.
\item[41:55] Yeah, I just put that in announcement in the chat then. So we also have some kickoff meeting so now we are still working on there on the organization but you will definitely hear from us as well. And in the in the chat I put the the link to our group.
\item[42:11] And, yeah, as I said that there would be more details. Soon and apparently there will be coordination both with this group, and with a long leave at high energy group.
\item[42:25] Okay, that's good, I realized this is early days, but, you know, on one hand, it's fantastic that everyone is welcoming everyone to join the groups. On the other hand, you know we can't all attend 20 Snowmass meetings every week. So, it would be great if somehow there was maybe, you know, of course it depends that there be a significant group of people are interested in a topic, but given the topic it'd be nice if there was sort of a home for it and maybe other groups are sort of, you know, supplementary or something. but if there could be sort of a central gathering area under one of the groups is sort of the primary group for that topic, that would be very helpful.
\item[43:10] Yeah, I agree I mean it's especially like topics like articles and all these things has a lot of common interest and so I, I think this world will self organize into materializing, there will be some centralized you know because because basically this is the same set of people.
\item[43:31] I think there will be that will will try to facilitate that and make it happen as much as possible, as well. Okay, there's also I think official liaisons between the different four tiers, so maybe one of their tasks could be that they make it clear when something is happening, about a given topic. And then they're, they're sort of the the connection between for the people who are interested in cross connection things.
\item[44:03] But again, I think you said is right it's early days so we're still trying to learn the most efficient thing I think we're we're all keep in mind that we don't like too many meetings. Yeah, exactly. I mean my read right now is that seems like EF oh nine very much sees the longer the particles and you know phases and things like that kind of thing and sort of a central topic in their group, whereas it seems a little bit on the side of this group of 10 and also Rp. Six, but yeah like that that's just my very early impression just from talking to people like for us so this is something that we've agreed within the, the energy frontiers that were our concern is the dark matter interpretation of those topics, and the four probably wasn't very clear in the previous presentation but what we will be.
\item[44:55] We'll talk more about the map searches.
\item[44:59] Because they are about stable connection and then whenever one wants to talk about the dark matter interpretation, then that's where the LLP searches will be considered in fo eight or nine, and then we talk about interpretations here.
\item[45:14] Okay, okay yeah that kind of guidance is really helpful. Yeah.
\item[45:19] Yeah, I mean how you want it to make yeah I just wanted to underscore what you both said that within the energy frontier topical groups we made an effort to understand the overlaps across topical groups, realizing really that while cross communication between topical groups are very necessary, the ownership of a given topic has to be within one topical group right because this is where the discussions will be nucleated and minimizes everybody to go to all meetings. So for instance, top EF or nine topical group will take the leadership, or the ownership, I should say, for the long live particles But clearly, people have to communicate with ef 10 and therefore eight more as Katrina underlined that interpretations is what the dark matter group will do but most of the work with similar analysis ideas and major overlaps will be part of your 409 and similarly for other you know model based interpretations are here for eight and generic is here for nine, and I see Jim you has resigned after me who was one of the Year for nine conveners so he can also elaborate on it, but yes we are very mindful of the, you know, efficient use of time for people because this is a very short and compressed process as well.
\item[46:50] Yeah, this is time, like together with Sam oma Angelica, we are yeah for nigh commoners. I think we were really asked menarche said we really value a time of the, of the community and our activity in the next few bi weekly meetings, you know, probably won't be organized the way our topics. So, when we come to lonely particles, we certainly will coordinate with the many relevant groups, including year after year for eight year for two and also the real processes groups to make sure other interested parties can join at Yeah, I'm just saying we have, we will assign seams for each bi weekly meetings.
\vspace{0.2cm}
\hrule
\newpage
\item[47:47] So there is another comment from Jordan.
\item[47:51] That is, we could try to encourage for run to pm SSM scans from Atlas and CMS to incorporate more dark matter constraints in their interpretations. This is something that.
\item[48:05] Another collaboration should should certainly do, I think. Right. Yeah, I think we kind of have a facilitating role there was because we are in the collaboration or member of the people that are interested here are also in the collaborations, but if we, if we plan for saying we would like to have this particular scenario, because it's one of the ones that we want to think about more carefully in Snowmass projection than this is something that would encourage.
\item[48:37] And so he's saying that they could have some extra guidance, the calibration could have extra guidance to benefit your plans.
\item[48:48] I tend to agree that if we have to.
\item[48:53] It will take a bit of time I think before we, we are into a state of saying that we definitely want to go in this direction.
\item[49:02] But the, it will come, I think.
\vspace{0.2cm}
\hrule
\vspace{0.2cm}
\item[49:09] Yeah So Carlos also raised the topic of, you know, including CP violation in the in the interaction between dark metal and stellar model I think it is has not been explicitly discussed here and. But yes, certainly that's especially Carlos he wanted to say a few words about it, I know somebody commented that this is more appropriate to EFL nine, however I seen that it's very difficult to draw the boundaries between this group.
\item[49:40] And the groups, so I said that the issue of separation is important, I encounter that class and impact on the way you can say on direct protection has also impact on of course, electric dipole moments, and class impact on collider physics, of course, and then, so I seen that they were motivated the possibility that the disappear relation interactions if you consider that these are sector related to the one that generate were the Genesis.
\item[50:11] And I believe that the sub facility that one should take into account. That's all what I wanted to say.
\item[50:20] Yeah, so it sounds like it's another topic across the different topical groups, again you know we'll be, we'll be focusing on its connection was it was documentary interpretations.
\item[50:32] But
\item[50:42] any additional comments.
\vspace{0.2cm}
\hrule
\newpage
\item[50:52] So looks like we are already 10 minutes past the time and, you know, perhaps we should we should close and, and just another reminder you know please fill in the Google Form and join our slack channels email list, so give us your input, and thanks everybody for for participating and, you know, we're working together in the future. Thank you very much, and our uncluttering up for taking care of this. Yeah, I thought in general, when he was connecting what would be nice if there is smarter participation of the experimentalists in the group.
\item[51:34] Then, because most of the people who speak up are theories
\item[51:41] and very shy theorists.
\vspace{0.2cm}
\hrule
\vspace{0.2cm}
\item[51:46] Yes, so I think on on the, on that sense it's, it's not that the experimentalists are not noisy but there's also some discussion that needs to go on within the experiments before they can speak up on behalf of one experiments, I think that's why it is a bit more, that they are more responsible.
\item[52:07] Okay, very good. Sorry for being too
\item[52:11] big for yourself.
\item[52:13] Okay thanks everybody. Yeah, thanks again to Antonio via for taking notes and closed captioning.
\item[52:20] Thank you very much. Oh, thank you. Thank you. Bye bye.
\vspace{0.2cm}
\hrule
\end{description}

%% file: misc/app_autocraption_ef_human_notetaking.tex
\SetLabelAlign{leftmargin}{\llap{#1\enspace}}
\SetLabelAlign{rightmargin}{\rlap{\hspace{\columnwidth}\enspace#1}}

\begin{center}
\Large Notes by a Physicist\footnote{Non-professional captioner}
\end{center}

\begin{description}[align=rightmargin,labelsep=0pt,leftmargin=*,font=\color{gray}]
\item[AD] comment on something both mentioned: multiplets. Good emphasis on higgsino and wino (from SUSY) but there are other multiplets that are less motivated from a model-building point of view - talking about minimal DM idea of what is a multiplet. There are other possibilities. It would be good to have benchmarks and studies on those scenarios.
\item[?] Agree. Minimal DM on LTW’s slide as well. Do you know what is the main difference? I think the cross section is different… 
\item[AD] The cross section is different. The main signal is (once beyond the triplet) you have objects with charge 2 or so on. (Live long enough to tell the charge?) There are well-tempered scenarios with mixtures of doublets and triplets. We have SUSY in the back of our mind, but [you can go beyond that with different spins etc.].
\item[JZ] even with doublet and triplet, if one changes the spin of DM, the pT distributions change (important for disappearing tracks).
\item[TT] generalizing spins is really important and simple to do
\item[TH] There are studies with higher gauge representations that typically lead to higher dimensional operators with several higgs fields. There are many more parameters…
\item[LTW] but they’re coupling to SM gauge bosons is determined by simple gauge interactions [which makes the production coupling simple to deal with].
\item[SK] Dirac or Majorana neutralinos (another way of extending beyond minimal SUSY). The interplay with DD experiments is also different for these options.
\item[LTW] in simple mediator models, Dirac vs Majorana can have a qualtitative difference (e.g. for t- channel).
\item[LTW] the important thing for snowmass is interesting benchmarks/goalposts. It’s usually not a complete exploration of model space.
\vspace{0.2cm}
\hrule
\newpage\phantom{blabla}\newpage
\item[SH] just trying to probe what is proper to be added and not. Often DM models are organized according to whether DM is charged under the SM or portals/hidden valleys. Most of the topics covered today are in these categories. Is this the full list of ways to tunnel into/probe into the dark sector?
\item[LTW] we wonder about that too. Do you have any alternatives?
\item[SH]  is it ok to throw out random thoughts?
\item[LTW] more importantly, write down [here] and tell us your interests.
\item[SH] example: go beyond gauge-invariant local operators. Are there global reasons for the DM to couple to SM? This would be in a different category. Concretely, DM couples to SM due to some shared anomaly, inducing an interaction in the low-energy effective theory. Some work in progress.
\vspace{0.2cm}
\hrule
\newpage
\item[AD] go beyond the minimal DM scenarios when you have co-annihilation with other particles, inspired by SUSY again. Eg., a singlet and a colored particle (that doesn’t have to be a gluino). 
\item[LTW] slides already had some SUSY-motivated co-annihilation scenarios. These give us additional lines to think about.
\item[TT] also models with non-tree-level mediation between SM and DM. 
\item[TR] Has there been any work done on loop-mediation?
\item[TT] Yes!  Check out e.g https://arxiv.org/abs/1506.01408. What I had in mind was models wher ethere is no tree-level mediation. Everything is intrinsically loop level, which is somewhat different than what LTW discussed [with loop-level corrections to tree-level] … There can be a very different momentum dependence in the loop-mediated models. [This could lead to pheno that is rather different].
\vspace{0.2cm}
\hrule
\newpage\phantom{blabla}\newpage
\item[MH] ratio of theorists to experimentalists in the WG?
\item[CD, LTW] a pretty good split for now
\vspace{0.2cm}
\hrule
\newpage
\item[YDT] On slide 23, several of  the dark photon constraints and projections are not updated. See https://gitlab.com/philten/darkcast for compilation of updates. I can volunteer to help keep the future summary plots updated (ytsai@fnal.gov)
\item[CD] will also have to discuss with Rare\&Precision frontier subgroup of dark sectors (RP06)
\item[SG (convenor)] dark sector studies should be done in cooperation with [the other groups].
\item[LTW] [a very important topic for a future meeting]
\item[PI] also interested/involved
\item[?] Long-lived particles are officially managed by EF09 but there is a lot of overlap and we should communicate. Here, the focus is the DM connection with the LLPs.
\item[SG] Following up on the previous discussion, this is the group that Mike Williams and I are convening: \url{https://snowmass21.org/rare/dark}. There will be coordination with this group. We will also announce a kickoff meeting soon.
\item[?] [to reduce the number of meetings that need to be attended, could there be a central gathering area under one of the groups as a home for a particular topic?]
\item[?] [the hope is that this will self-organize to an extent, assisted by the conveners and official liaisons between frontiers. It’s early days so we’re still trying to learn the most efficient thing to do.]
\item[JF] EF09 sees LLPs as a central topic, whereas it is on the side of this and other groups. It would be good to understand where searches for LLP go, especially FASER/Codex-b/MATHUSLA/etc.  It seems these are more central to EF09, but also RF06.  It would be nice if a “home” subgroup could be identified so that the relevant activity can be focused and not fall between the cracks (and also so the interested parties don’t have to attend 10 Snowmass meetings each week!)
\item[CD] the concern here is the DM interpretation of these topics.
\item[JF] that guidance is helpful.
\item[MN] we’ve tried to understand the overlap, realizing that cross communication is necessary but the ownership of a given topic has to be within one topical group. This is where the discussions will be nucleated and this minimizes the number of overall meetings. For LLP, EF09 takes ownership, but clearly communication with EF10 and EF08 is needed. Interpretations is what the DM group will do. Similarly for other model-based interpretations (EF08). We are very mindful of the efficient use of time.
\item[ZL] (convener of EF09): we really value the time of the community, and in the next few biweekly meetings we will probably organize by topic. When it comes to LLP, we will coordinate with the other groups to make sure the other interested parties can join. We will assign themes for each meeting.
\vspace{0.2cm}
\hrule
\newpage
\item[GS] note that one thing we can try to encourage now for the Run-2 pMSSM scans from ATLAS and CMS is to incorporate more DM constraints in the SUSY reinterpretations.
\item[LTW] this is something that the collaborations should certainly do.
\item[CD] Snowmass can have a facilitating role there. If we plan that we would like to have a particular scenario, this is something we [should] encourage [from the collaborations]. The collaborations should have some extra guidance to benefit future plans. It will take a bit of time before we are able to say we want to go in a particular direction, but it will come.
\vspace{0.2cm}
\hrule
\vspace{0.2cm}
\item[CW] One of the topics that I don’t know if it has  been discussed (I was late) is the possibility of CP violation in the interaction between the Dark sector and the Standard one. This could be well motivated if the Dark sector has anything to do with the origin of the baryon asymmetry. This can have important consequences for direct detection, for instance.
\item[LTW] CP violation has not been discussed yet. Indeed an important topic to discuss.
\item[CW] somebody commented that this is more appropriate to EF09, but I think it is very difficult [to set boundaries on this]. This has an impact on what you can say about DD and also on collider physics. It is well-motivated and should be taken into account.
\item[LT] another topic that can be cross-group, cross-connection with DM interpretation. 
\vspace{0.2cm}
\hrule
\newpage
\item[LTW] Reminder: please fill in the google form to give us your input, thanks for participating and we’ll work together in the future. 
\vspace{0.2cm}
\hrule
\vspace{0.2cm}
\item[AD] More participation of experimentalists would be good, most of the people are theorists
\item[CD] experimentalists are organizing within the collaboration, may be early to speak about certain topics
\vspace{0.2cm}
\hrule
\end{description}